\documentclass[pdflatex,sn-mathphys-num]{sn-jnl}

\usepackage{graphicx}%
\usepackage{multirow}%
\usepackage{amsmath,amssymb,amsfonts}%
\usepackage{amsthm}%
\usepackage{mathrsfs}%
\usepackage[title]{appendix}%
\usepackage{xcolor}%
\usepackage{textcomp}%
\usepackage{manyfoot}%
\usepackage{booktabs}%
\usepackage{algorithm}%
\usepackage{algorithmicx}%
\usepackage{algpseudocode}%
\usepackage{listings}%

\usepackage{mathptmx}          
\usepackage[T1]{fontenc}       
\usepackage[utf8]{inputenc}    

\theoremstyle{thmstyleone}%
\theoremstyle{thmstyletwo}%

\theoremstyle{thmstylethree}%

\newcommand{\hxmt}{\emph{Insight}-HXMT}

\newcommand{\nicer}{\emph{NICER}}

\begin{document}

\title[Reverberation lags viewed in hard X-rays from an accreting stellar-mass black hole]{Reverberation lags viewed in hard X-rays from an accreting stellar-mass black hole}


\author*[1]{\fnm{Bei} \sur{You}}\email{youbei@whu.edu.cn}

\author*[2,3]{\fnm{Wei} \sur{Yu}}\email{yuwei@ihep.ac.cn}

\author[4]{\fnm{Adam} \sur{Ingram}}

\author[5]{\fnm{Barbara} \sur{De Marco}}

\author[2]{\fnm{Jin-Lu} \sur{Qu}}

\author[6,1]{\fnm{Zong-Hong} \sur{Zhu}}

\author[3]{\fnm{Andrea} \sur{Santangelo}}

\author[1]{\fnm{Sai-En} \sur{Xu}}

\affil[1]{\orgdiv{Department of Astronomy}, \orgname{School of Physics and Technology, Wuhan University}, \orgaddress{\city{Wuhan}, \postcode{430072}, \country{China}}}

\affil[2]{\orgdiv{Key Laboratory of Particle Astrophysics}, \orgname{Institute of High Energy Physics, Chinese Academy of Sciences}, \orgaddress{\city{Beijing}, \postcode{100049}, \country{China}}}

\affil[3]{\orgdiv{Institut für Astronomie und Astrophysik, Kepler Center for Astro and Particle Physics}, \orgname{Eberhard Karls Universität}, \orgaddress{\street{Sand 1}, \postcode{72076}, \city{Tübingen}, \country{Germany}}}

\affil[4]{\orgdiv{School of Mathematics, Statistics, and Physics}, \orgname{Newcastle University}, \orgaddress{\city{Newcastle upon Tyne}, \postcode{NE1 7RU}, \country{United Kingdom}}}

\affil[5]{\orgdiv{Departament de Física, EEBE}, \orgname{Universitat Politècnica de Catalunya}, \orgaddress{\street{Av. Eduard Maristany 16}, \postcode{08019}, \city{Barcelona}, \country{Spain}}}

\affil[6]{\orgdiv{Department of Astronomy}, \orgname{Beijing Normal University}, \orgaddress{\city{Beijing}, \postcode{100875}, \country{China}}}


\abstract{\textbf{The X-ray-emitting corona near a black hole (BH) is too small to be directly imaged, but the rapid variability is used to infer the geometry by measuring time lags caused by coronal X-rays reflecting off the disk, known as reverberation lags. Though reverberation lags have previously been detected for some supermassive BHs in active galactic nuclei (AGNs), detecting them from stellar-mass BHs poses much greater challenges due to their size being over a million times smaller. Previous measurements of reverberation lags for stellar-mass BHs were limited to energies below 10 keV. Here, we report the detection of the Compton hump reverberation, peaking at about 30 keV, from an X-ray binary. The accompanying detection of an iron line feature at about 6.4 keV confirms the X-ray reverberation scenario and provides strong evidence that accretion flows in AGNs and X-ray binaries are governed by an ubiquitous process.}}

\maketitle

\section*{Introduction}

A black hole X-ray binary (BHXRB) emission from the central region around the BH when it is triggered into an outburst\cite{lasota2001}. The low energy (soft) photons (below a few keV) from the outer disk are inverse Compton scattered within the inner hot flow and/or jet near the BH, accounting for the observed high energy (hard) power-law emission, up to a cutoff energy of about 100 keV. In this work, we refer to the X-ray Compton scattering source as the corona. 
During a typical outburst, an X-ray binary transitions from the corona-dominated \textit{hard state} to the disk-dominated \textit{soft state} and back again, before fading once more into quiescence.
The X-ray emission is highly variable on time-scales of milliseconds to tens of seconds \cite{wilkinson2009accretion}, so X-ray timing analysis is a very powerful tool for probing the innermost flow around accreting BHs \cite{uttley2002,uttley2011,cassatella2012joint,uttley2014,mendez2022coupling,2022NatAs...6..577M, uttley2025large}.
When long timescales of seconds to tens of seconds are isolated by Fourier analysis (i.e., considering Fourier frequencies lower than 1 Hz), variations in higher energy (harder) X-rays are seen to lag behind variations in lower energy (softer) X-rays. This is the so-called `hard lag'. Such a low-frequency hard lag is thought to originate from fluctuations in the mass accretion rate, which propagate from the outer disk to the central corona \cite{2001MNRAS.327..799K,2006MNRAS.367..801A}. If the corona is large, a significant hard lag will also be contributed by the extra light travel time associated with higher energy photons that have experienced more Compton scattering events \cite{kylafis2008jet}. The propagation of fluctuations in the disk arises due to turbulence induced by the magneto-rotational instability \cite{balbus1998instability}. However, our understanding of this process has historically been limited by past measurements of the hard lag usually only being available in the $<$ 30 keV energy range, whereas the corona radiates strongly up to energies $> 100$ keV. \cite{nowak1999,2001MNRAS.327..799K,demarco2015,de2017evolution,kara2019}.

The Comptonized photons out of the corona can partially scatter back to the outer cold disk, where they are reprocessed, producing the observed reflection features from 0.5 keV to hundreds of keV \cite{fabian2010x,garcia2013,dauser2016relativistic}, which include the relativistic iron K emission line at about 6.4 keV and the Compton hump at 20-70 keV \cite{aaz2004,you2021}. This disk reflection component is expected to lag behind the emission from a variable corona due to light-crossing delays, producing X-ray reverberation features \cite{fabian2009,uttley2014}. 

Whereas hard lags dominate at low Fourier frequencies, reverberation lags can be seen at high Fourier frequencies, i.e., in the fastest variability. The iron K reverberation lag and the Compton hump lag were first discovered in active galactic nuclei (AGNs) \cite{zoghbi2014observations,kara2015}. The Fourier frequency of about $10^{-4}$ Hz at which these lags were discovered corresponds to about $10^2$ Hz for an X-ray binary, since all variability timescales are thought to scale linearly with BH mass \cite{mchardy2006active,arevalo2006spectral}. Searching for high-frequency lags in BHXRBs has, therefore, recently been a major focus in X-ray timing studies \cite{wilkinson2009accretion,uttley2011}. However, mass scaling makes detection challenging, since the 1 ks lag observed for AGNs with a $10^7~M_\odot$ BH corresponds to only 1 ms for an X-ray binary with a $10~M_\odot$. Nonetheless, iron K reverberation lags have been observed \cite{de2017evolution,kara2019,2022ApJ...930...18W} using high throughput soft X-ray detectors, but the Compton hump lag has not been found yet despite extensive searches.

The frontier for X-ray timing is thus to extend the analysis to higher energy bands, which is vital for gaining a complete understanding of the accretion flow close to the BHs and the associated physical processes.
This is now possible with the Hard X-ray Modulation Telescope (\hxmt{}) \cite{zhang2014}. As one of the three instruments in \hxmt{}, the high energy X-ray Telescope (HE) can detect X-ray photons in the 20-250 keV energy range. 
The effective area at $E=20-100$ keV is about 0.5 $\rm m^2$, and decreases to about $0.1 \rm m^2$ at $E=200$ keV.
In the past four years, \hxmt{} has made significant contributions to the spectral and timing analysis of BHXRBs.
In the latest BHXRB catalog of \hxmt, MAXI J1820+070 is one of the brightest sources, which makes it the most promising candidate to probe the timing and spectral aspects of BHXRB in the high energy band, ensuring high X-ray counts and excellent statistics \cite{you2021,you2023}.

\begin{figure} 
	\centering
	\includegraphics[width=\textwidth]{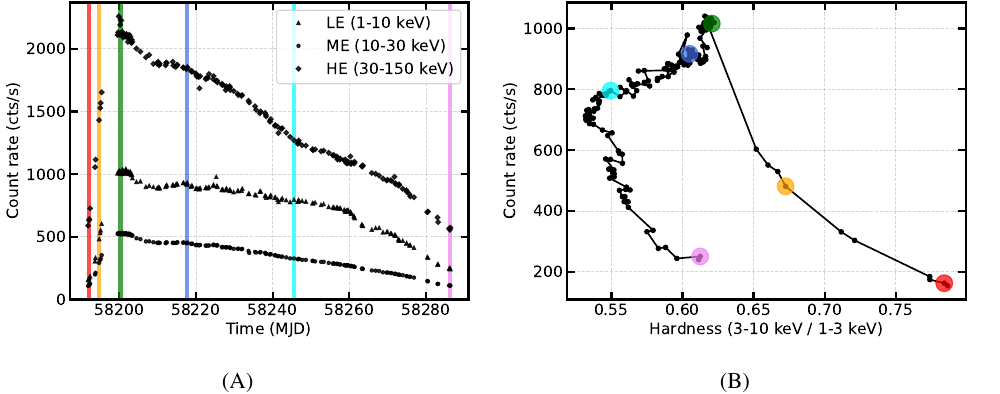}
    \caption{\textbf{\hxmt{} lightcurves and corresponding hardness-intensity diagram.}
		(\textbf{A}) \hxmt{} lightcurves (in units of counts per second) of MAXI J1820+070 in HE (27-250 keV), ME (10-30 keV) and LE (1-10 keV) band, in the rising hard state during the outburst ranging from MJD 58192 to 58286. (\textbf{B}) The corresponding hardness-intensity diagram, defined as the total 1–10 keV count rate (in units of counts per second) versus the ratio of hard (3–10 keV) to soft (1–3 keV) count rates. The central times of the six epochs of interest are shown as colored points, corresponding to MJD = 58192, 58194, 58200, 58217, 58245, and 58286.}
	\label{hxmt_lightcurve} 
\end{figure}

In this work, we focus on the timing analysis of high energy radiation from the corona, and mainly concentrate on the \hxmt{} observations that catch the source in the rising hard state (MJD = 58192-58286). Fig.~\ref{hxmt_lightcurve}A shows the \hxmt{} light curves from the low-energy (LE, from 1 to 10 keV), medium-energy (ME, from 10 to 30 keV), and high-energy (HE, from 27 to 150 keV) instruments (see `Methods' for a description of the data reduction). The hardness-intensity (HID) diagram, defined as the total 1–10 keV count rate (in units of counts per second) versus the ratio of hard (3–10 keV) to soft (1–3 keV) count rates, is plotted in Fig.~\ref{hxmt_lightcurve}B. 
Owing to the enormous size of the dataset, we describe only the timing results of six epochs of the total \hxmt{} observations. These epochs were selected for their lengthy observations, each capturing a different characteristic phase of the hard state. The first three epochs cover the period between the first observation and the light curve peak, and the remaining three epochs cover the period of the light curve decay. These six observational epochs, marked in Fig.~\ref{hxmt_lightcurve}, correspond to ObsIDs P0114661{001, 002, 003, 016, 040, 070}, taken on 2018-03-14 (MJD 58192), 2018-03-16 (MJD 58194), 2018-03-22 (MJD 58200), 2018-04-09 (MJD 58217), 2018-05-07 (MJD 58245), and 2018-06-17 (MJD 58286). The respective exposure times for these observations are 6.4 ks, 34 ks, 33 ks, 6.3 ks, 3.7 ks, and 2.3 ks.

\section*{Results and Discussion}

\subsection*{Frequency-dependent time lags}
 
\begin{figure} 
	\centering
	\includegraphics[width=\textwidth]{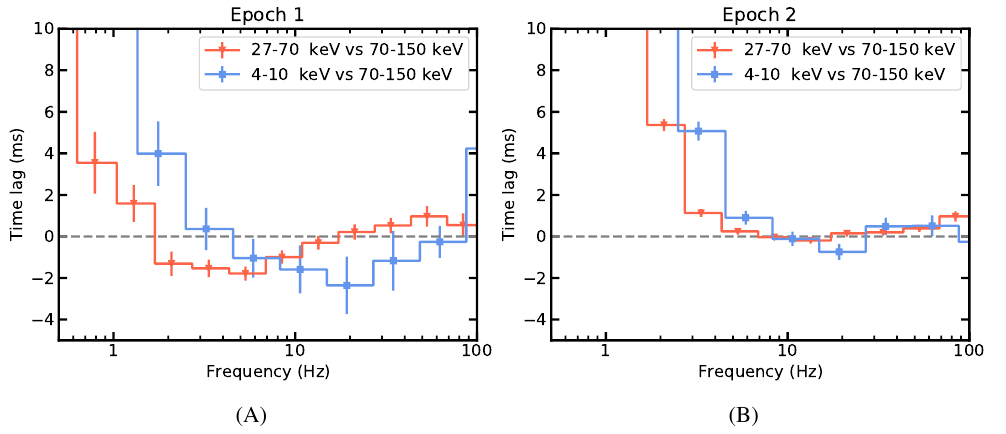}
    \caption{\textbf{Lag-frequency spectra of MAXI J1820+070.}
(\textbf{A}) Lag-frequency spectrum for Epoch 1; (\textbf{B}) Lag-frequency spectrum for Epoch 2. In each panel, the blue line shows the lag between the 4--10 keV and 70--150 keV bands, and the red line shows the lag between the 27--70 keV and 70--150 keV bands. A negative lag indicates that the softer band lags behind the harder band. Error bars indicate 1$\sigma$ uncertainties.}

	\label{lag_f} 
\end{figure}

The spectral fits to the \hxmt{} spectra of MAXI J1820+070 show that the disk reflection flux is comparable with the corona Comptonization flux below 70 keV at early epochs of the rising hard state (see Supplementary Fig.~1 in \cite{you2021}). In contrast, the emission at 70-150 keV is dominated by the cutoff power law of the corona Comptonization, making it a suitable reference band for probing X-ray reverberation between the corona and the disk. 
We perform frequency-resolved timing analysis for the six epochs and measure the frequency-dependent time lag between the 70-150 keV and the 27-70 keV emission, as well as between the 70-150 keV and the 4-10 keV emission. Fig.~\ref{lag_f} shows the frequency-dependent time lags for the first two epochs of \hxmt{}. In the first epoch (left panel), the 70-150 keV emission lags behind the two softer bands (4-10 keV and 27-70 keV) at low frequencies of $<$1 Hz (hard lag), while it leads both softer bands at high frequencies (soft lag). 

Such high-frequency soft lags detected in the response of both the 27-70 keV and the 4-10 keV emission to the variable hard (70-150 keV) X-ray flux could be attributed to a delayed response (reverberation) of the disc component to variations of the corona, given the presence of the Compton hump and the iron K line significantly contributing to the flux in each of the two bands. This will be further justified below by studying the energy-dependent lag at the frequency range where these soft lags are detected.

We also note that, at the highest frequencies, the lag crosses zero and becomes hard once more. This is consistent with the expectations of phase-wrapping, whereby the measured lag changes sign when the true (intrinsic) time delay, \(\tau_{\mathrm{int}}\), exceeds the variability timescale associated with the Fourier frequency (analogous to car wheels appearing to run backwards in a film due to the rotation rate of the wheels being greater than the frame rate of the camera). We find that this characteristic phase-wrapping frequency is approximately an order of magnitude higher for the iron K reverberation than for the Compton hump reverberation. Similar differences in reverberation frequencies between the Fe-K and Compton hump bands have not been reported in AGNs.

In the simplest case, the characteristic phase-wrapping frequency \(f_{\mathrm{wrap}}\) is related to the intrinsic delay via \(f_{\mathrm{wrap}} = 1/(2\,\tau_{\mathrm{int}})\). Adopting a black hole mass of \(10\,M_\odot\), and using the observed phase-wrapping frequencies in epoch~1, we estimate that the iron line reverberation originates from \(R_{\mathrm{Fe}} \approx 51\,r_g\) and the Compton hump from \(R_{\mathrm{CH}} \approx 145\,r_g\) (where \( r_g = GM/c^2 \)). These distances should be interpreted as weighted means over all possible light-travel paths of the Comptonised photons to the disc, and can be significantly larger than the minimum light-travel distance set by the X-ray source height and the inner disc radius \citep{wilkins2013, mahmoud2019reverberation,demarco2021}.

These estimates are broadly consistent with the scenario in which the ionization state of the accretion disc depends on radius: closer to the black hole, the disc is more heavily irradiated and therefore more highly ionized, enhancing the iron line strength, whereas the strength of the Compton hump is largely independent of ionization state \cite{garcia2013}. This implies that the iron line emission originates from a more compact region than the Compton hump, naturally producing a higher characteristic phase-wrapping frequency.
Additional discussion of the frequency-dependent behavior of the soft lag between the 27--70 keV and 4--10 keV bands is available in the Methods section.

In the second epoch, just two days after the first epoch, the high-frequency soft lags of both the 27-70 keV and the 4-10 keV emission are highly suppressed (see the right panel of Fig.~\ref{lag_f}), while the low-frequency hard lags show a substantial increase. This combination suggests that there must have been a dramatic evolution of the accretion flow on a daily timescale. To investigate the origin of this lag evolution, we further examine the energy-dependent lags below.

\subsection*{High-frequency lag spectrum}

\begin{figure} 
	\centering
	\includegraphics[width=0.8\textwidth]{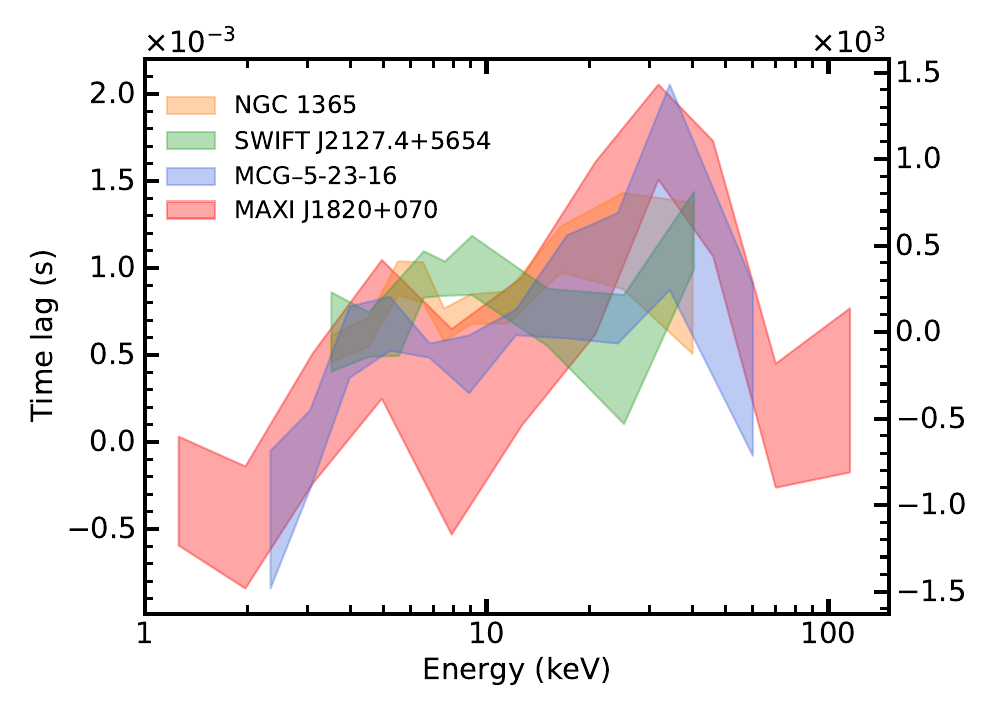}
	\caption{\textbf{Comparison of high-frequency lag-energy spectra between AGNs and XRB}. The left y-axis displays the lag measurements for XRB MAXI J1820+070 during epoch 1, using a reference band of 1-10 keV, within the 2-60 Hz frequency range where significant high-energy soft lags were observed as depicted in Fig.~\ref{lag_f}. The right y-axis shows the high-frequency lag-energy spectra for three AGNs: MCG–5-23-16, SWIFT J2127.4+5654, and NGC 1365. These AGNs exhibit detected iron K and Compton hump reverberation lags, highlighting the similarities in the reverberation processes across different scales of black hole systems \cite{zoghbi2014observations, kara2015}. The shaded region indicates the 1$\sigma$ uncertainty range.}
	\label{lag_e} 
\end{figure}

The observation at the first epoch is the only one in which the soft lags between 70-150 keV and 27-70 keV are detected. The soft lags fall within the frequency range of approximately 2-60 Hz. We, therefore, estimate energy-dependent lags in this frequency range for the first epoch by measuring the lag between several narrow `subject bands' and one common `reference band'. We adopt the 1–10 keV band as the reference band, with photons from the band of interest removed to avoid noise correlation. The resulting lag-energy spectrum is plotted in Fig.~\ref{lag_e} (red shaded region). In the energy band below 10 keV, the lag spectrum displays a broad feature that aligns with the characteristics of a relativistically broadened iron line. 
Above 10 keV, the lag rises to a maximum at about 30 keV and decreases at higher energies. Therefore, the overall lag spectrum in the 10-100 keV range resembles the Compton hump in the energy spectrum of the disk reflection. 
The presence of features that resemble the iron K and the Compton hump in the lag spectrum confirm the X-ray reverberation to be the physical origin of the observed high-frequency soft lag \cite{zoghbi2014observations,kara2015}.

It is generally thought that AGNs are mass-scaled-up XRBs \cite{mchardy2006active,arevalo2006spectral}. The iron K and Compton hump reverberation lags have also been detected in three AGNs to date, i.e., MCG–5-23-16, SWIFT J2127.4+5654 and NGC 1365 \cite{zoghbi2014observations,kara2015}. We plot the lag spectra of these three AGNs in Fig.~\ref{lag_e} to compare it with the reverberation lags of the XRB. The AGN lag-energy spectra are directly taken from \citet{zoghbi2014observations} and \citet{kara2015} without any modification of the original data. Despite their diverse shapes, the AGN lag spectra show substantial overlap with the XRB lag spectrum, particularly in the case of MCG-5-23-16.
Notably, the masses of these three AGNs are all on the order of $10^7 M_{\odot}$ \cite{wandel1986observational,marconi2003relation,malizia2008first,ponti2012caixa,emmanoulopoulos2014general}, differing by six orders of magnitude from MAXI J1820+070. This mass difference well matches the observed difference in lag amplitude, consistent with the expected linear scaling of reverberation lag with black hole mass ($\tau \propto M_{\rm BH}$). This provides the strongest evidence of the mass scaling of reverberation lag amplitude between stellar-mass black holes and AGNs, offering observational support for a ubiquitous process in the accretion flows surrounding both stellar-mass and supermassive black holes.

It is also important to note that the mass scaling of reverberation lag is observed only in epoch 1. The disappearance of the reverberation signal after epoch 1 complicates a simple picture of mass scaling across black hole systems. This may reflect the fact that AGNs evolve on much longer timescales than stellar-mass XRBs, making it difficult to observe comparable lag spectral evolution within the available observational windows. It may also be due to the limited number of studies on high-energy lag spectra in AGNs, which hinders the identification of AGN analogues to the XRB behavior observed in epochs 2–6. Further investigation of AGN lag–energy spectra at higher energies and frequencies will be essential to determine whether the behavior observed in epochs 2–6 also occurs in supermassive black holes.

\subsection*{Evolution of high-frequency lag spectrum}

The timing analysis of \nicer{} data showed the presence of soft lags in low energy bands (2-5 keV vs. 0.5-1 keV), likely arising from X-ray reverberation, which could be detected throughout all the epochs of the outburst of the source before it transitioned to the soft state \cite{demarco2021}. However, by extending the time lag analysis to much higher energy bands, i.e., the 70-150 keV v.s. 27-70 keV, we find that the reverberation lags are highly suppressed after the first epoch (see the right panel of Fig.~\ref{lag_f}).

\begin{figure} 
    \centering
    \includegraphics[width=\textwidth]{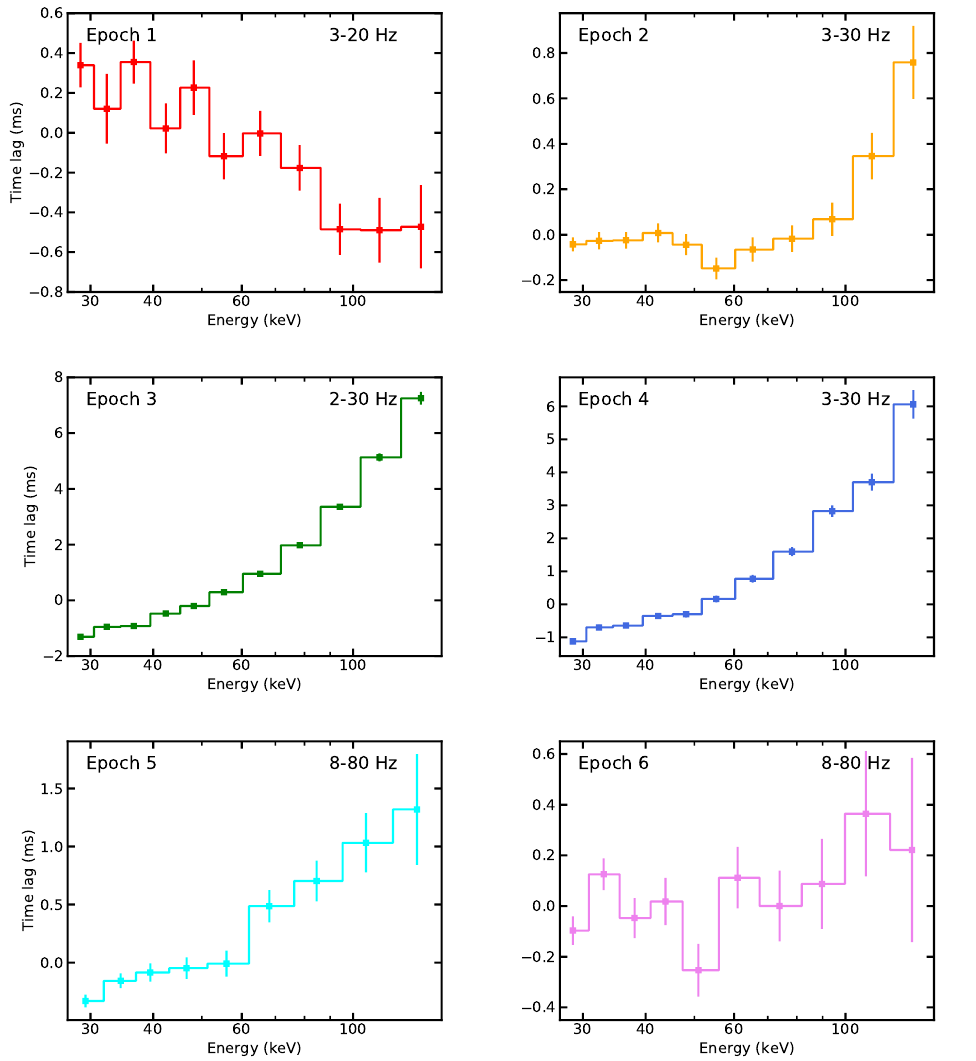}
    \caption{\textbf{Evolution of the lag-energy spectra for the six observation epochs.}
    The frequency range used to produce the lag-energy spectra corresponds to the soft lag region identified in the lag-frequency spectra from the \nicer{} data. The reference band is selected as 27-150 keV. The color of each observation matches that in Fig.~\ref{hxmt_lightcurve}. Error bars indicate 1$\sigma$ uncertainties.}
    \label{lag_e_evo} 
\end{figure}

To further investigate the physical origin of high-energy soft lags and their evolution, as illustrated in Fig.~\ref{lag_f}, we then focus on analyzing the observations by the HE instrument. 
The frequency range of the soft lag, estimated using the lag-frequency spectra from quasi-simultaneous \nicer{} data (within $\pm$ 1 day), is employed to derive the lag-energy spectra for the hard X-rays and to analyze their evolution during the rising hard state. 
Fig.~\ref{lag_e_evo} presents the lag-energy spectra at six epochs between MJD 58192 and 58286. The lag is measured between each small energy bin and a broad reference band (27-150 keV), with the bin of interest removed to avoid noise correlation. It is evident that at the onset of the outburst, high-energy photons reach the observer first, followed by lower-energy reflected photons, leading to a decrease in lag with increasing energy, ranging from approximately +0.4 ms to -0.5 ms (epoch1). The shape of the lag energy spectrum follows the tail of the Compton hump as shown in Fig.~\ref{lag_e}.
As the flux from the source increases rapidly (epoch 2), the lag-energy spectra exhibit significant changes, pivoting around 60 keV so that high-energy photons from the Comptonization tail (E $>$ 70 keV) arrive later than the low-energy photons arising from reflection. 

As the source flux reaches its peak during the rising hard state (e.g., epoch 3), the spectrum becomes predominantly governed by hard lags, which linearly increase with energy (in logarithm) between 30-150 keV, ranging from about -1 ms to 
about +7 ms. It can be seen that the high-energy photons lag the low-energy photons by almost 8 ms at this peak flux. During the flux-decreasing phase after leaving the peak, the lags start to decrease. Compared with the initial phase (epoch1), where the low-energy photons lag the high-energy photons by $< 1$ ms, it becomes evident that the lag behavior evolves significantly throughout the hard state. 
We also find that at the end of the hard state, when the X-ray count rate is quite low, the lags weakly depend on the photon energy, and their amplitude is highly reduced,  similar to early \hxmt{} observations during the outburst. 

From the high-frequency lag-energy spectra, it is apparent that with the increase in source brightness, the soft lags associated with reverberation are rapidly obscured by strengthening hard lags. Such hard lags have been extensively observed and studied in BHXRBs and are widely believed to result from inward-propagating perturbations in the mass accretion rate \cite{lyubarskii1997flicker,2001MNRAS.327..799K,2006MNRAS.367..801A}. These propagation-induced lags typically dominate at low Fourier frequencies and decrease with increasing frequency. Previous studies have shown that the frequency dependence of hard lags can often be described by a power-law trend \cite{1999ApJ...510..874N,2000A&A...357L..17P}. Due to this frequency dependence, hard lags generally have minimal impact on the high-frequency lag spectrum—this, however, holds only when their amplitude remains modest. The presence of strong hard lags in high-frequency spectra suggests that they may represent an extension of low-frequency propagation-induced lags into higher frequencies. To test this hypothesis, we next turn to an analysis of the low-frequency lag–energy spectra, which are typically dominated by such propagation-induced hard lags.

\subsection*{Low-frequency lag spectrum and its evolution}

\begin{figure} 
	\centering
	\includegraphics[width=\textwidth]{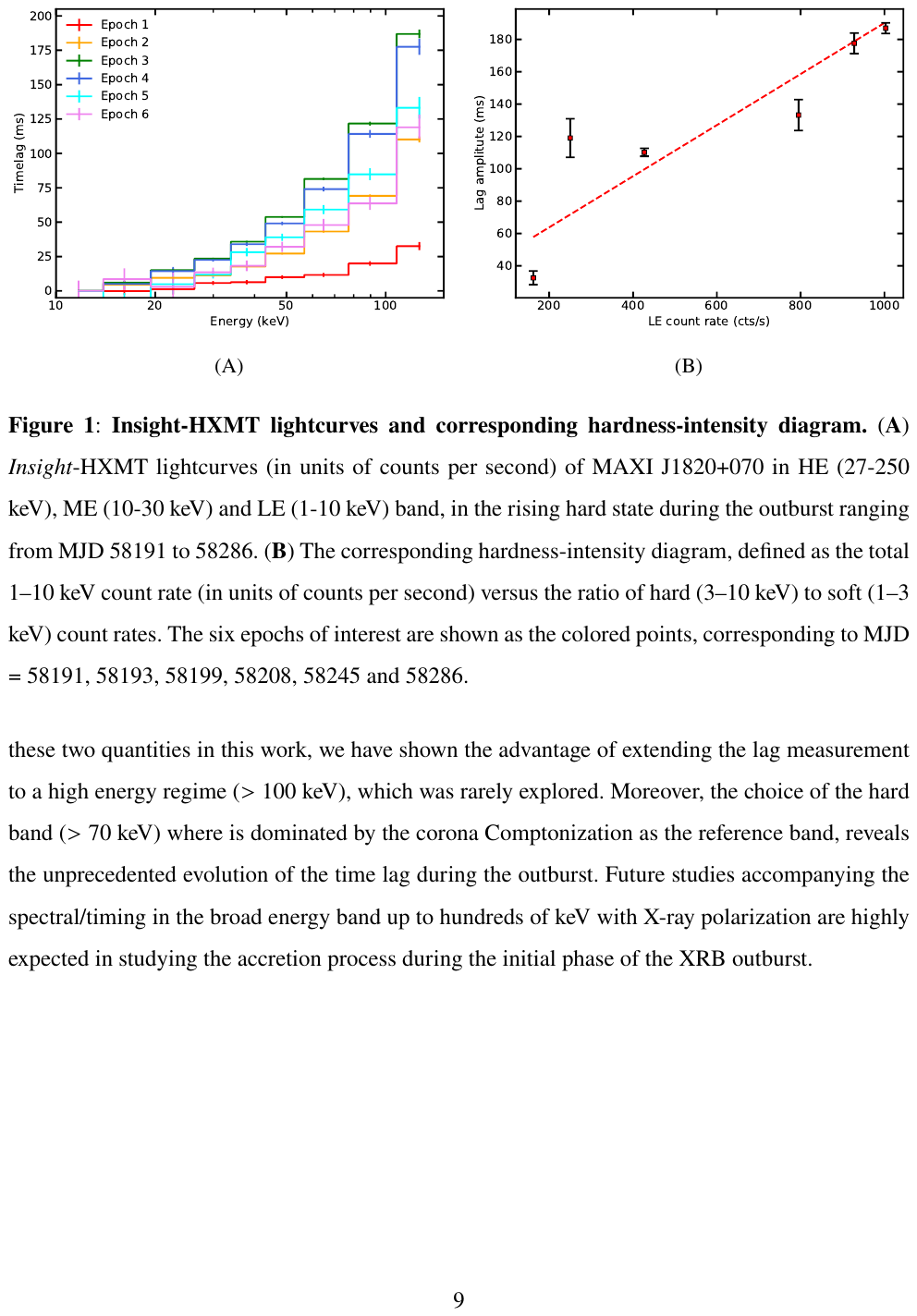}
    \caption{\textbf{Low-frequency lag-energy spectra of MAXI J1820+070.}
    \textbf{(A)} The 0.1–1 Hz lag-energy spectra for the six epochs. The reference band is chosen to be 1-10 keV. The spectra cover the energy range of the ME and HE instruments (10-150 keV). The lags are shifted so that the lowest-energy bin is set to zero. The color of each observation matches the colors used in Fig.~\ref{hxmt_lightcurve}. 
    \textbf{(B)} The maximum amplitude of the hard lags as a function of the LE count rates. The red dashed line shows the best-fitting linear relation. Error bars indicate 1$\sigma$ uncertainties.}

	\label{lag_e_evo_low_f} 
\end{figure}

We measured the inter-band time delays by averaging over the frequencies (0.1-1 Hz) where the hard lag is predominant, as shown in Fig.~\ref{lag_f}. The lag is measured between each energy bin and a broad reference band (1-10 keV). The derived lag spectrum and its evolution for different epochs are plotted in Fig.~\ref{lag_e_evo_low_f}. Lag--energy spectra were also reported by \citet{wang2020evolution} in the 1--10 Hz range, but our choice of the 0.1--1 Hz band better isolates the dominant hard lag component and, in addition, includes the earliest stages of the outburst that were not covered in their study.

At epoch 1, the time-lag shows only a weak dependence on photon energy. By epoch 2, however, the low-frequency hard lags increase rapidly-by a factor of about four-and become more pronounced toward the high-energy end (left panel of Fig.~\ref{lag_e_evo_low_f}). As the amplitude grows, the hard lags begin to extend into higher frequencies and gradually suppress the soft lags (Fig.~\ref{lag_e_evo}; see also Fig.~\ref{lag_f}), providing a plausible explanation for the disappearance of high-energy reverberation lags after epoch 1.
After epoch 3, i.e., the peak of the count rate, the time-lag spectra soften with a drop of the lag in the high-energy end. Notably, despite the dramatic evolution of the hard lags between epochs 1 and 2, their corresponding energy spectra show no significant change. This suggests that the amplitude of the hard lag may be more strongly correlated with luminosity rather than spectral hardness. For example, the luminosity in epochs 2 and 6 is approximately the same, while the spectral hardness differs considerably, yet the lag spectra appear very similar. To further explore this trend, we measured the maximum lag amplitude in each epoch and found that the hard lag amplitude shows a positive correlation with the count rate (see right panel of Fig.~\ref{lag_e_evo_low_f}). Notably, the evolution of hard lags is primarily pronounced at higher energies and becomes increasingly significant as the energy rises. No obvious evolution of hard lags is observed below 30 keV.

The hard lags are thought to be predominantly due to perturbations of the mass accretion rate propagating inward within the hot corona \cite{lyubarskii1997flicker,2001MNRAS.327..799K,2006MNRAS.367..801A}.
In this scenario, perturbations in the accretion rate originate at large radii and travel inward through the hot corona. These perturbations modulate the X-ray emission at different energies and radii, leading to energy-dependent time delays. To translate this into a quantitative model, the X-ray emission at each energy is assumed to follow a radial emissivity profile of the form \( \propto r^{-\gamma(E)} \), where \( \gamma(E) \) is the energy-dependent emissivity index. A steeper index at higher energies (i.e., \( \gamma(E_{\rm h}) > \gamma(E_{\rm s}) \)) implies that higher-energy photons predominantly originate from smaller radii. This naturally leads to hard lags, since fluctuations arrive at the inner regions later in time \cite{2012MNRAS.419.2369I}. A recent study by \citet{kawamura2023maxi} applied a model based on propagating fluctuations to \hxmt{} data of MAXI J1820+070. While their original model could not explain the lag behavior above 10~keV, an updated version incorporating spectral pivoting succeeded up to 50~keV. Beyond this range, discrepancies remain, likely due to the absence of a physical description for the pivoting process.

In our analysis, we adopted a well-established model, \texttt{PROPFLUC} \cite{2011MNRAS.415.2323I,2012MNRAS.419.2369I,2013MNRAS.434.1476I,2014MNRAS.440.2882R,2016MNRAS.462.4078R}, to interpret the observed hard lag evolution in Fig.~\ref{lag_e_evo_low_f}. This model simulates how the observed lag spectrum changes in response to two key physical parameters: (1) the emissivity index difference between energy bands, \( \Delta \gamma = \gamma(E_{\rm h}) - \gamma(E_{\rm s}) \), and (2) the outer radius of the corona, \( r_{\rm o} \), which sets the spatial scale of the fluctuations. Our simulations suggest that the dramatic increase and subsequent softening of the hard lag can be explained by changes in one or both of these parameters. Specifically, when the lag variation is driven by changes in coronal size, the corona appears to expand from approximately \(17\,R_{\rm g}\) at epoch~1 to \(50\,R_{\rm g}\) at epoch~3, and then contract to \( 32\,R_{\rm g} \) by epoch~6. These results are consistent with earlier spectral-timing studies. Additional details on the simulations can be found in the Methods section.

\section*{Summary}

In this work, the X-ray reverberation mapping measurement was extended to a broadest energy range of 1-150 keV. We report the detection of a Compton hump reverberation feature from the X-ray binary MAXI J1820+070, along with an accompanying iron line feature that supports the interpretation of X-ray reverberation mapping as the underlying mechanism. Moreover, the lag-energy spectrum shows similarity to that observed in AGNs when scaled by mass, providing strong evidence that the accretion flows in AGNs and X-ray binaries are governed by a ubiquitous process.

We also find a clear difference in the reverberation time scales of the iron K emission line and the Compton hump. This may represent observational evidence, from a timing perspective, of an ionization gradient in the accretion disk, which highlights the potential value of incorporating such gradients in future spectral modeling efforts.

On longer time scales, the lag-energy spectrum reveals the rapid evolution of the corona during the initial phase of the outburst, suggesting that the coronal properties at this stage differ significantly from those in later stages. This finding highlights the necessity of employing wide-field monitors to capture the very onset of black hole binary outbursts—an accretion phase that has been overlooked. Early detections are crucial for enabling high-quality follow-up spectral, timing, and polarimetric observations with next-generation missions, such as the enhanced X-ray Timing and Polarimetry (eXTP) \cite{extp} and Athena\cite{athena}. These advancements will enable exploration of the physics governing extreme accretion processes during the nascent stages of outburst activity.

\clearpage

\section*{Methods}

\subsection*{Data reduction}

The Insight-HXMT satellite is equipped with three primary instruments designed to cover a broad X-ray energy range: the High Energy X-ray Telescope (HE; 20–250 keV), the Medium Energy X-ray Telescope (ME; 5–30 keV), and the Low Energy X-ray Telescope (LE; 1–15 keV).

We processed the observational data using version 2.05 of the HXMT Data Analysis Software (HXMTDAS). Standard screening procedures were applied to exclude periods with unstable pointing or elevated background, adopting the following criteria: (1) pointing offset less than $0.05^\circ$; (2) source elevation above the Earth limb exceeding $6^\circ$; and (3) geomagnetic cutoff rigidity greater than 6. Instrumental background was modeled using an empirical correlation approach involving detectors with small field-of-view (FoV) and their corresponding blind detectors. The scaling factor, derived from the ratio of active to blind detectors, was used to estimate background contributions. This background estimation technique has been validated using observations of blank-sky regions.

\subsection*{Time-lag measurements}

X-ray reverberation mapping is a very powerful tool for probing the innermost regions of BHXRBs and AGNs \cite{zoghbi2010,zoghbi2014observations,kara2015,kara2019,demarco2021,2022ApJ...930...18W}. In this study, we adopt a frequency-domain analysis method rooted in Fourier transform techniques \cite{nowak1999,vaughan2003,uttley2014} to quantify the time lags between light curves in different energy bands. We begin by binning light curves from selected energy bands at a time resolution of 0.001 seconds. The discrete Fourier transform (DFT) is applied to each light curve, and the cross-spectrum is computed as the product of the Fourier transform of one energy band and the complex conjugate of another. This provides access to the relative phase information between signals. To improve signal-to-noise, the cross-spectrum is averaged over logarithmically spaced frequency intervals. The phase lag at each frequency bin is then converted into a time lag by dividing by $2\pi f$, where $f$ denotes the central frequency of the bin. In our convention, a positive lag indicates that variations in the hard X-ray band lag behind those in the soft X-ray band.

\subsection*{On the energy dependence of the characteristic phase wrapping frequency}\label{secA1}

In the first observing epoch, we find that the frequency at which phase wrapping begins, $\nu_0$, is approximately an order of magnitude higher for the $4-10$ vs $70-150$ keV lags than it is for the $27-70$ vs $70-150$ keV lags. Although the soft lag amplitude itself does not directly indicate the reverberation lag, $\tau$, due to dilution (i.e. both energy bands include contributions from direct and reflected components), it has been shown that the frequency at which the soft lag transitions to a hard lag due to phase wrapping is unaffected by dilution \cite{wilkins2013,uttley2014,mizumoto2018,demarco2021}. This implies that $\nu_0 \propto 1/\tau$ should be independent of the energy bands used to measure the time lag. However, this inference is only true if the emergent reflection spectrum is the same for all disk radii. In reality, the ionization parameter is a strong function of radius due to the radial dependencies of both the irradiating flux and the disk density \cite{Svoboda2012,Ingram2019a,Shreeram2020}. The ionization parameter strongly influences the strength of the iron line, but not the Compton hump \cite{garcia2013}, meaning that the iron line essentially originates from a smaller region than the Compton hump does, and thus its associated reverberation lag is smaller, leading to $\nu_0$ being higher.

To test this interpretation, we create a new model within the X-ray reverberation mapping software package \textsc{reltrans} \cite{Ingram2019a,Mastroserio2021}. The existing \textsc{reltrans} models output time lag as a function of energy for a given Fourier frequency range. The new model that we create here, \texttt{reltransCpF}, outputs the time lag between two user-defined energy bands as a function of Fourier frequency.

The underlying model is the same as \texttt{reltransDCp} \cite{Mastroserio2021}. The corona is assumed to be a point-like `lamppost' source a height $h$ above the black hole that illuminates a flat disk with inner and outer radii $r_{\rm in}$ and $r_{\rm out}$ that is observed from an inclination angle $i$. Other physical parameters are the black hole spin $a$, the photon index $\Gamma$, the ionization parameter $\log\xi$, the iron abundance relative to solar $A_{\rm Fe}$, the disk density $\log n_e$, the coronal electron temperature $kT_e$, the black hole mass $M$, and the \texttt{boost} parameter, which regulates the fraction of the observed flux that is reflected. The coronal spectrum is modeled with the thermal Comptonisation code \texttt{nthComp} \cite{Zdziarski1996} and the restframe reflection spectrum is modeled using \texttt{xillverCp} \cite{garcia2013}.

Reverberation lags are calculated self-consistently in the assumed geometry, whereas hard lags are created by introducing variability to the photon index $\Gamma(t)$ \cite{Mastroserio2018}. The variability in $\Gamma$ at a Fourier frequency $\nu$ is governed by the parameters $\gamma(\nu)$ and $\phi_{AB}(\nu)$, where the former is the variability amplitude relative to that of the normalization, and the latter is the phase lag between the $\Gamma$ variability and that of the normalization. In the usual \textsc{reltrans} models, the user specifies a different value of $\gamma$ and $\phi_{AB}$ for each frequency range considered, and these are left as free parameters. For the new frequency-dependent model, we must instead parameterize the frequency dependence of $\gamma(\nu)$ and $\phi_{AB}(\nu)$. We employ the formulae
\begin{equation}
\phi_{\rm AB}(\nu) = \phi_{\rm AB,0} ~\nu^{-s_{\rm AB}},~~~~~~~~\gamma(\nu) = \gamma_0 ~\nu^{-s_{\gamma}}.
\end{equation}
where $\phi_{\rm AB,0}$, $s_{\rm AB}$, $\gamma_0$ and $s_{\gamma}$ are model parameters.

\begin{figure} 
	\centering
	\includegraphics[width=\textwidth]{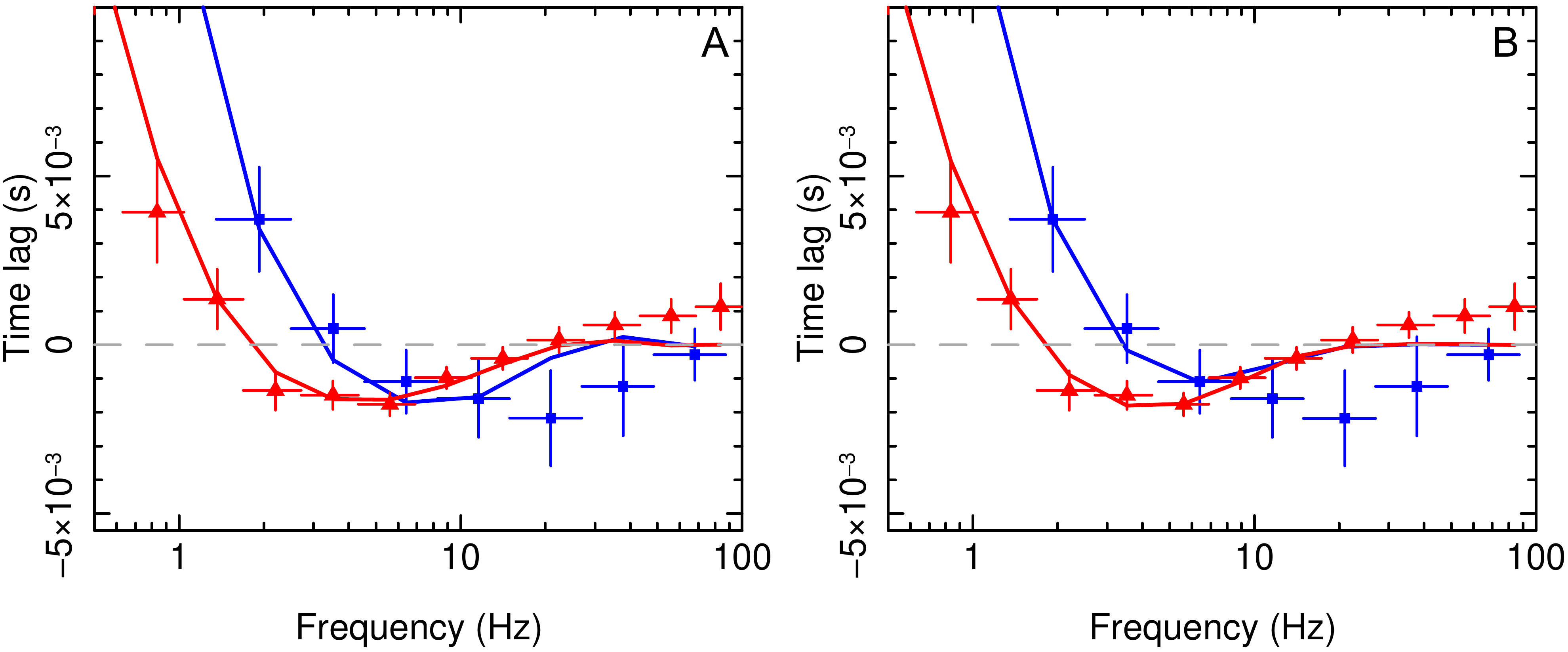}
	\caption{\textbf{Example fits to the epoch 1 lag spectrum with a reverberation model.} We use a new model in the \textsc{reltrans} package to model the epoch 1 lag vs frequency spectra for $4-10$ keV vs $70-150$ keV (blue squares) and $27-70$ keV vs $70-150$ keV (red triangles). \textbf{(A)} The model accounts for a radial ionization gradient, enabling phase wrapping to begin at a different frequency for the two lag spectra. \textbf{(B)} The model assumes a constant disk ionization parameter, meaning that phase wrapping always begins at the same frequency.}
	\label{fig:reltrans} 
\end{figure}

We attempt preliminary fits of this new model to the epoch 1 \hxmt{} data. Since interstellar absorption is not important for the high photon energies we are considering, the lag spectrum is completely independent of the instrument response matrix \cite{Ingram2019}, and thus we do not account for the instrument response matrices. Fig.~\ref{fig:reltrans}A shows the result of fitting \texttt{reltransCpF} when accounting for the radial dependence of $\log\xi(r)$ by employing a grid of 20 radial zones and assuming a constant disk density. We see that for this model, the frequency at which phase wrapping begins is indeed different for the two different lag spectra: the zero-crossing point is about 20 Hz and 30 Hz for the Compton hump and iron line lags, respectively. We note that this difference is smaller than what is seen in the data, but the model is consistent with the data within uncertainties. In contrast, Fig.~\ref{fig:reltrans}B shows the result of fitting the model under the assumption that $\log\xi$ is independent of radius. We see that, as expected, the characteristic phase wrapping frequency is the same for both lag spectra. The parameters of the best fitting model are not believable (e.g. $i\approx 7^\circ$, $M\approx 34~M_\odot$), but after trying many parameter combinations, we reliably find that $\nu_0$ can only depend on the energy bands used if a radial $\log\xi$ gradient is employed. We additionally tried different assumptions for the density profile, first using the zone A Shakura-Sunyaev \cite{Shakura1973} density profile, $n_e \propto r^{3/2} [ 1 - \sqrt{r_{\rm in}/r}]^{-2}$, and then a simpler $n_e \propto r^{3/2}$ profile. In both cases, the best-fitting lag spectra were very similar to the constant density case, albeit with slightly different (but still not especialy believable) best fitting parameters. In future, we will attempt a formal fit of the epoch 1 data with the \texttt{reltransCpF} model.

\subsection*{Simulation of Low-Frequency Hard Lags with the PROPFLUC Model}

The \texttt{PROPFLUC} model incorporates fluctuations in the mass accretion rate that propagate through the hot accretion flow, as well as the precession of the entire hot flow induced by frame-dragging effects near the black hole. In this framework, accretion rate fluctuations originate at all radii within the hot flow and travel inward toward the black hole, giving rise to a broadband noise component in the power spectrum. The characteristic timescale of this variability is governed by the local viscous timescale within the flow \cite{1997MNRAS.292..679L,2001MNRAS.321..759C,2006MNRAS.367..801A}. Due to the inward propagation of these fluctuations, the variability in the emission from different annuli of the flow is temporally cross-correlated, with time lags corresponding to their respective propagation delays.

For the simulations, we adopt the standard parameter configuration of the \texttt{PROPFLUC} model. The surface-density profile is normalized with $\Sigma_{0}=10$, characterized by profile indices $\kappa = 3$, $\lambda = 0.9$, and $\zeta = 0$. The bending-wave radius is fixed at $r_{\rm bw}=7$, and the fractional variability per radial decade is set to $F_{\rm var}=0.3$. The black hole mass is taken to be $M = 10~M_\odot$, with a dimensionless spin parameter $a=0.998$. The flow is divided into $N_{\rm dec} = 35$ rings per decade in radius. The inner radius is fixed at $r_{\rm i}=4.5~R_{\rm g}$, while the outer radius $r_{\rm o}$ is treated as a free parameter to explore the dependence of the propagating fluctuation lags on corona size.

\begin{figure} 
    \centering
    \includegraphics[width=0.7\textwidth]{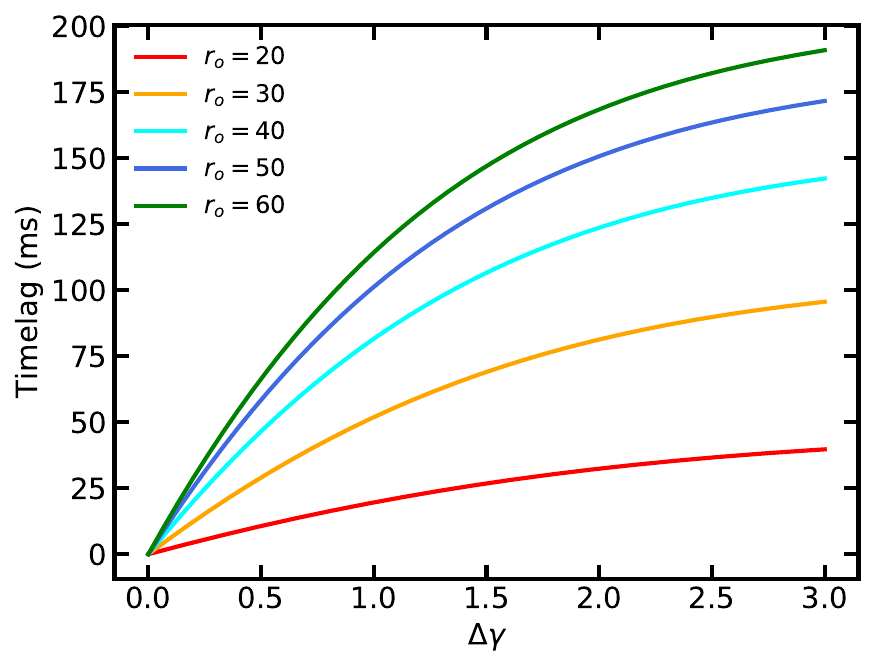}
    \caption{\textbf{Simulated low-frequency (0.1–1 Hz) lag spectra using the \texttt{PROPFLUC} model.}
        The low-frequency (0.1–1 Hz) lag spectra simulated using the propagating mass accretion rate fluctuations model \texttt{PROPFLUC}. Most model parameters were fixed, with only the emissivity index and the outer radius of the corona varying.}
    \label{simulation} 
\end{figure}

We adopt an emission index of $\gamma (E_{\rm s}) = 2.0$ for the reference (soft) energy band, where the emission index describes how rapidly photons of a given energy decay with radius across the accretion flow. Typically, higher-energy photons exhibit steeper radial decay (i.e., larger $\gamma$), while lower-energy photons are more extended in radius. The parameter $\Delta \gamma = \gamma (E_{\rm h}) - \gamma (E_{\rm s})$ quantifies the difference in these radial emissivity profiles between the hard and soft bands. A positive $\Delta \gamma$ naturally leads to propagation lags between energy bands, since fluctuations travel inward through regions emitting differentially in energy. The greater the $\Delta \gamma$, the larger the expected time lag, due to a steeper energy gradient in the spatial distribution of emission. In contrast, when $\Delta \gamma = 0$, both energy bands share the same emissivity profile, and no lag due to propagation would be observed. The results of the simulations are shown in  Fig.~\ref{simulation}.

Based on our fluctuation model simulations, the weak energy-dependence of the hard lag at epoch 1 (left panel of Fig.~\ref{lag_e_evo_low_f}) can be explained in two physical scenarios:
(1) a small difference in the emissivity profile between the soft and hard bands $\gamma (E_{\rm h}) \simeq \gamma (E_{\rm s})$, meaning that both soft and hard photons are emitted across similar radial regions, and thus propagation lags are minimal. 
(2) a large difference in the emissivity profile with $\gamma (E_{\rm h}) \gg \gamma (E_{\rm s})$, but a spatially compact corona (e.g., $r_{\rm o} \leq 20 Rg$), limiting the radial propagation range and thus suppressing observable lags.

As for the first scenario, the hardening of the hard lag spectrum since epoch 1 would require increases in the emissivity index difference $\Delta \gamma = \gamma (E_{\rm h}) - \gamma (E_{\rm s})$ and/or the size of the corona $r_{\rm o}$. The softening of the hard lag spectrum after epoch 3 (the peak of the count rate) may correspond to a decrease in $\Delta \gamma$ and/or $r_{\rm o}$.

As for the latter scenario, the hardening and softening of the hard lag spectrum mainly correspond to an increase and a decrease of $r_{\rm o}$, respectively. Indeed, the large $\Delta \gamma$ implies a weak dependence of the hard lag on this parameter.
Therefore, a softening of the hard lag spectrum after epoch 3 would suggest a contraction of the corona. This is interesting as this discovery is consistent with the conclusions by the \nicer{} spectral/timing analysis \cite{kara2019,demarco2021,2022ApJ...930...18W}. We then attempt to make a quantitative estimate of the coronal size. Since the dependence of the emissivity profile index $\gamma$ on energy is unknown, we assume an extreme case for $\Delta\gamma = 6$ and estimate the coronal size using the maximum observed lag amplitude. It should be noted that this assumption may lead to an underestimate of the actual coronal size. The results suggest that the coronal size increases from $17 R_{\rm g}$ at epoch 1 to $50 R_{\rm g}$ at epoch 3, and then gradually decreases to $32 R_{\rm g}$ by epoch 6.

The corona size and its emissivity profile are important properties in picturing the accretion flow near the BH during the initial phase of the outburst. Although it is hard to put constraints on these two quantities in this work, we have shown the advantage of extending the lag measurement to a high energy regime ($>$ 100 keV), which was rarely explored.  Future studies of broad spectral/timing analysis up to hundreds of keV, possibly combined with X-ray polarization (in the IXPE energy band), are expected to further our knowledge of the accretion process and its evolution during an outburst.

\subsection*{Additional Timing and Spectral Results}

\begin{figure} 
	\centering
	\includegraphics[width=\textwidth]{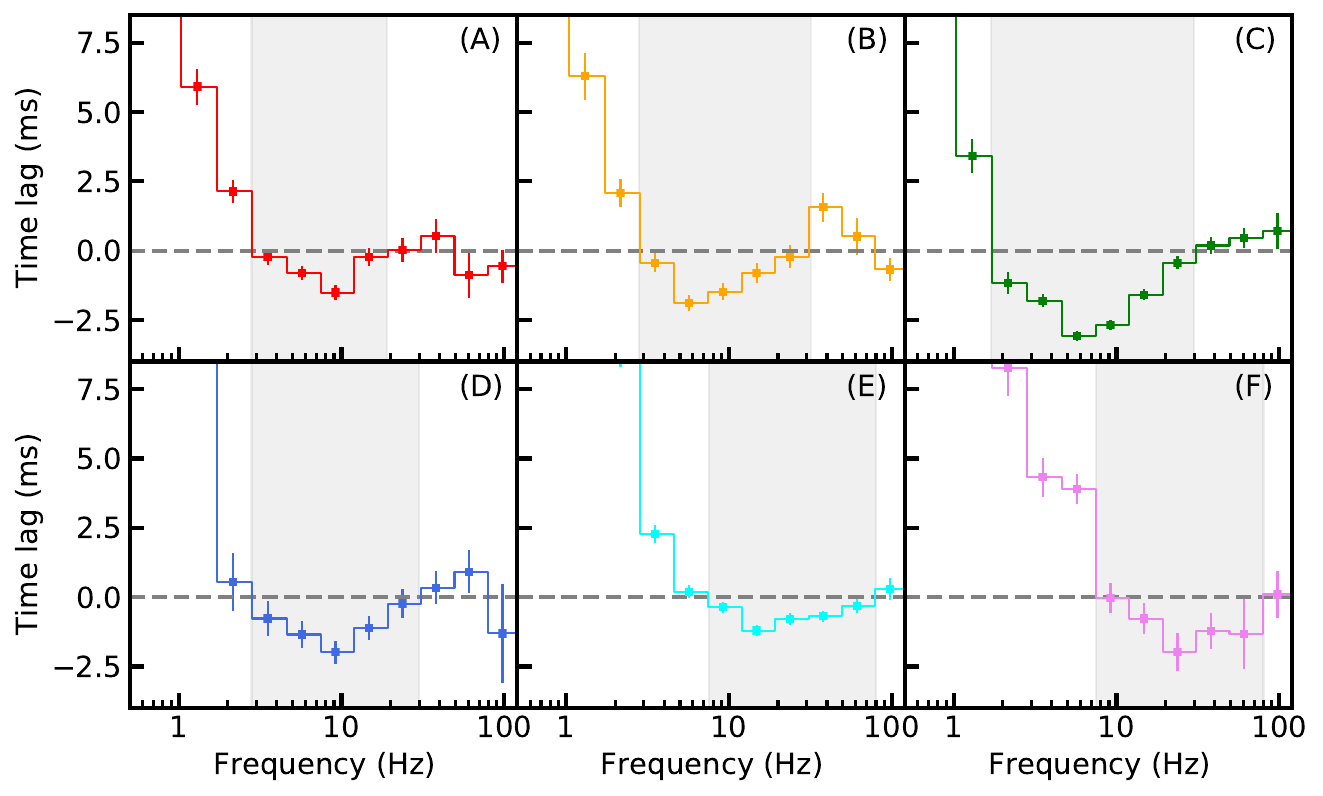}
	\caption{\textbf{The \nicer{} lag-frequency spectra measured between 0.5–1 keV and 2–5 keV.} (A)-(F) correspond to quasi-simultaneous \nicer{} observations for epochs 1-6, respectively. The shaded regions indicate where a soft reverberation lag is detected, with these frequency ranges employed to calculate the lag-energy spectra in Fig.~\ref{lag_e_evo}. Error bars indicate 1$\sigma$ uncertainties.}
	\label{nicer} 
\end{figure}

\begin{figure} 
    \centering
    \includegraphics[width=0.7\textwidth]{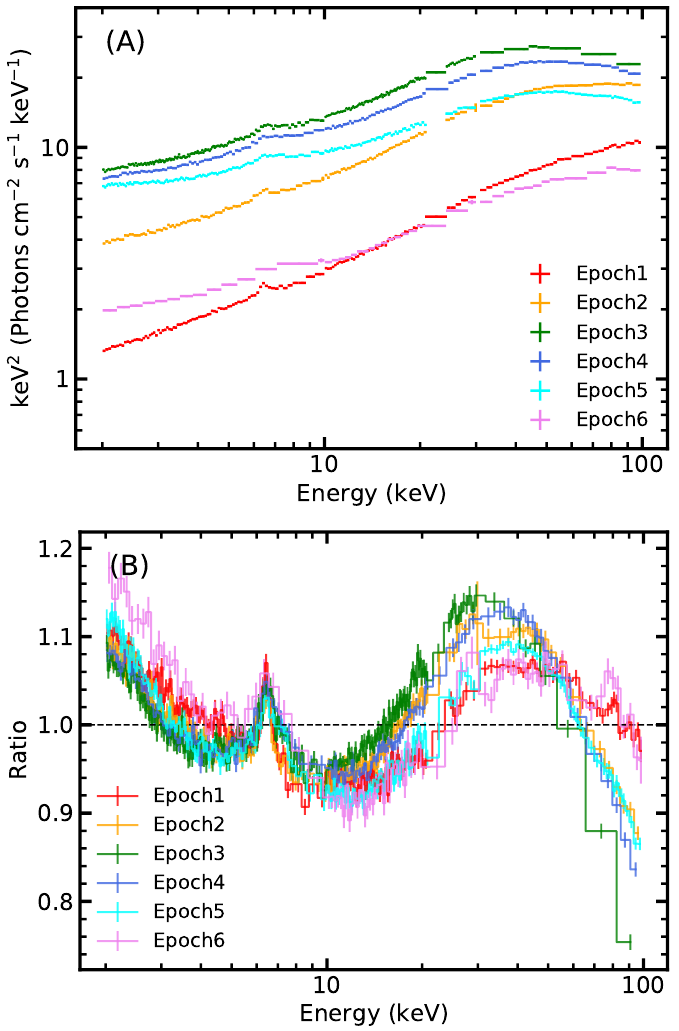}
    \caption{\textbf{\hxmt{} X-ray spectra from the selected observational epochs.}
        \textbf{(A)} \hxmt{} X-ray spectra of MAXI J1820+070 for six epochs. \textbf{(B)} Data-to-model ratios with respect to the best-fit \texttt{cutoffpl} model, highlighting the Fe--K line and Compton hump relative to the continuum. Error bars indicate 1$\sigma$ uncertainties.}
    \label{spec} 
\end{figure}

\begin{figure} 
    \centering
    \includegraphics[width=0.8\textwidth]{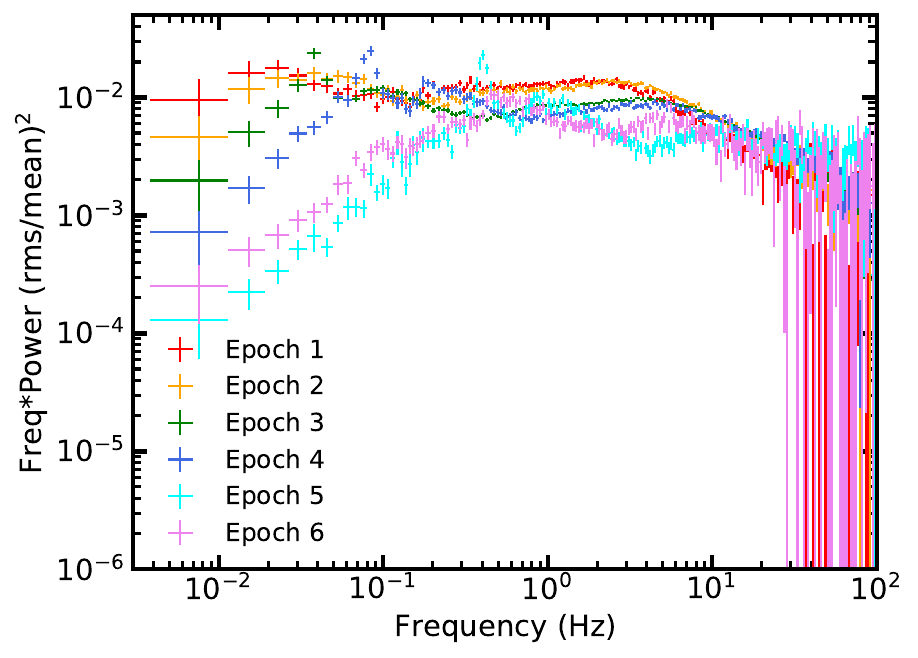}
    \caption{\textbf{Power spectra from the Compton hump band.} \hxmt{} power spectra extracted from the 27–70 keV band for the six epochs. Vertical error bars indicate 1$\sigma$ uncertainties and horizontal error bars indicate the frequency binning.}
    \label{pds} 
\end{figure}

The \nicer{} lag-frequency spectra measured between 0.5-1 keV and 2-5 keV are shown in Fig.~\ref{nicer}. The frequency ranges in which the soft lags are detected are used to generate the corresponding lag-energy spectra presented in Fig.~\ref{lag_e_evo}.
Representative X-ray spectra from the selected epochs, together with their data-to-model ratios relative to the best-fitting \texttt{cutoffpl} model, are shown in  Fig.~\ref{spec}. The Iron line and Compton hump inferred from these spectra align well with those seen in the lag–energy spectra, both in terms of energy range and relative strength. In addition, we also present the power spectra extracted from the Compton hump energy band (Fig.~\ref{pds}), which further illustrate its contribution to the observed variability.

\subsection*{Data availability}

All \hxmt{} data used in this work (Proposal ID: P0114661) are publicly available and can be downloaded from the \hxmt{} website (http://archive.hxmt.cn/proposal). The NICER datasets analysed during this study are available at NASA’s High Energy 
Astrophysics Science Archive Research Center (https://heasarc.gsfc.nasa.gov/FTP/nicer/data/obs/). The data generated in this study are publicly available at: \\https://github.com/MAXIJ1820/generated\_data. Source data are provided with this paper. 

\subsection*{Code availability}

The Insight-HXMT data reduction was performed using software available from the Insight-HXMT website (http://hxmten.ihep.ac.cn/). The time lag was performed with Stingray, a reliable Python library for X-ray timing analysis (see https://stingray.readthedocs.io/en/latest/index.html). The model PROPFLUC can be downloaded from the website: \\https://github.com/HEASARC/xspec\_localmodels/tree/master/propfluc. The new model presented here, \texttt{reltransCpF}, can be downloaded from https://github.com/reltrans/Youetal2025.

\section*{Acknowledgments}

B. Y. is supported by NSFC grants 12322307, 12361131579, 12273026; the National Program on Key Research
and Development Project 2021YFA0718500. Xiaomi Foundation / Xiaomi Young Talents Program. The data analysis in this paper have been done on the supercomputing system in the Supercomputing Center of Wuhan University.
W. Y. acknowledges support from the Sino-German (CSC–DAAD) Postdoc Scholarship Program (No. 57718047) and the Alexander von Humboldt Foundation.
A. I. acknowledges support from the Royal Society.
B.D.M. acknowledges support via a Ram\'on y Cajal Fellowship (RYC2018-025950-I), the Spanish MINECO grants PID2023-148661NB-I00, PID2022-136828NB-C44, and the AGAUR/Generalitat de Catalunya grant SGR-386/2021.
J.-L Q. acknowledges support from the grant U2031205.
Z.-H. Z. acknowledges support from the National Natural Science Foundation of China under Grants Nos. 12021003 and 12433001.

\section*{Author contributions:}
B.Y. initiated the project, specifically proposing data analysis and model interpretation, and took the lead in manuscript writing. W. Y. led the timing analysis and contributed to the text writing. A.I. led the interpretation of the Frequency-dependent time lags, and contributed to the writing of the text. B. D. contributed to the model discussion and the writing of the text. J.-L. Q., Z.-H. Z, A.S., and S.-E. X. contributed to the model discussion. All the authors joined in the modification of the text at all stages.

\section*{Competing interests:}
There are no competing interests to declare.

\section*{Figures:}

\clearpage 




\bibliography{sn-bibliography}

@Misc{methods,
  note = {Materials and methods are available as supplementary material},
}

@ARTICLE{lasota2001,
       author = {{Lasota}, Jean-Pierre},
        title = "{The disc instability model of dwarf novae and low-mass X-ray binary transients}",
      journal = {New Astronomy Reviews},
         year = 2001,
        month = jun,
       volume = {45},
       number = {7},
        pages = {449-508},
          doi = {10.1016/S1387-6473(01)00112-9},
archivePrefix = {arXiv},
       eprint = {astro-ph/0102072},
 primaryClass = {astro-ph},
       adsurl = {https://ui.adsabs.harvard.edu/abs/2001NewAR..45..449L}
}

@ARTICLE{you2023,
       author = {{You}, Bei and {Cao}, Xinwu and {Yan}, Zhen and {Hameury}, Jean-Marie and {Czerny}, Bozena and {Wu}, Yue and {Xia}, Tianyu and {Sikora}, Marek and {Zhang}, Shuang-Nan and {Du}, Pu and {Zycki}, Piotr T.},
        title = "{Observations of a black hole x-ray binary indicate formation of a magnetically arrested disk}",
      journal = {Science},
         year = 2023,
        month = sep,
       volume = {381},
       number = {6661},
        pages = {961-964},
          doi = {10.1126/science.abo4504},
archivePrefix = {arXiv},
       eprint = {2309.00200},
 primaryClass = {astro-ph.HE},
       adsurl = {https://ui.adsabs.harvard.edu/abs/2023Sci...381..961Y}
}

@ARTICLE{you2021,
       author = {{You}, Bei and {Tuo}, Yuoli and {Li}, Chengzhe and {Wang}, Wei and {Zhang}, Shuang-Nan and {Zhang}, Shu and {Ge}, Mingyu and {Luo}, Chong and {Liu}, Bifang and {Yuan}, Weimin and {Dai}, Zigao and {Liu}, Jifeng and {Qiao}, Erlin and {Jin}, Chichuan and {Liu}, Zhu and {Czerny}, Bozena and {Wu}, Qingwen and {Bu}, Qingcui and {Cai}, Ce and {Cao}, Xuelei and {Chang}, Zhi and {Chen}, Gang and {Chen}, Li and {Chen}, Tianxiang and {Chen}, Yibao and {Chen}, Yong and {Chen}, Yupeng and {Cui}, Wei and {Cui}, Weiwei and {Deng}, Jingkang and {Dong}, Yongwei and {Du}, Yuanyuan and {Fu}, Minxue and {Gao}, Guanhua and {Gao}, He and {Gao}, Min and {Gu}, Yudong and {Guan}, Ju and {Guo}, Chengcheng and {Han}, Dawei and {Huang}, Yue and {Huo}, Jia and {Jia}, Shumei and {Jiang}, Luhua and {Jiang}, Weichun and {Jin}, Jing and {Jin}, Yongjie and {Kong}, Lingda and {Li}, Bing and {Li}, Chengkui and {Li}, Gang and {Li}, Maoshun and {Li}, Tipei and {Li}, Wei and {Li}, Xian and {Li}, Xiaobo and {Li}, Xufang and {Li}, Yanguo and {Li}, Zhengwei and {Liang}, Xiaohua and {Liao}, Jinyuan and {Liu}, Congzhan and {Liu}, Guoqing and {Liu}, Hongwei and {Liu}, Xiaojing and {Liu}, Yinong and {Lu}, Bo and {Lu}, Fangjun and {Lu}, Xuefeng and {Luo}, Qi and {Luo}, Tao and {Ma}, Xiang and {Meng}, Bin and {Nang}, Yi and {Nie}, Jianyin and {Ou}, Ge and {Qu}, Jinlu and {Sai}, Na and {Shang}, Rencheng and {Song}, Liming and {Song}, Xinying and {Sun}, Liang and {Tan}, Ying and {Tao}, Lian and {Wang}, Chen and {Wang}, Guofeng and {Wang}, Juan and {Wang}, Lingjun and {Wang}, Wenshuai and {Wang}, Yusa and {Wen}, Xiangyang and {Wu}, Baiyang and {Wu}, Bobing and {Wu}, Mei and {Xiao}, Guangcheng and {Xiao}, Shuo and {Xiong}, Shaolin and {Xu}, Yupeng and {Yang}, Jiawei and {Yang}, Sheng and {Yang}, Yanji and {Yi}, Qibin and {Yin}, Qianqing and {You}, Yuan and {Zhang}, Aimei and {Zhang}, Chengmo and {Zhang}, Fan and {Zhang}, Hongmei and {Zhang}, Juan and {Zhang}, Tong and {Zhang}, Wanchang and {Zhang}, Wei and {Zhang}, Wenzhao and {Zhang}, Yi and {Zhang}, Yifei and {Zhang}, Yongjie and {Zhang}, Yue and {Zhang}, Zhao and {Zhang}, Ziliang and {Zhao}, Haisheng and {Zhao}, Xiaofan and {Zheng}, Shijie and {Zhou}, Dengke and {Zhou}, Jianfeng and {Zhu}, Yuxuan and {Zhu}, Yue},
        title = "{Insight-HXMT observations of jet-like corona in a black hole X-ray binary MAXI J1820+070}",
      journal = {Nature Communications},
         year = 2021,
        month = jan,
       volume = {12},
          eid = {1025},
        pages = {1025},
          doi = {10.1038/s41467-021-21169-5},
archivePrefix = {arXiv},
       eprint = {2102.07602},
 primaryClass = {astro-ph.HE},
       adsurl = {https://ui.adsabs.harvard.edu/abs/2021NatCo..12.1025Y}
}

@ARTICLE{ingram2019,
       author = {{Ingram}, Adam R. and {Motta}, Sara E.},
        title = "{A review of quasi-periodic oscillations from black hole X-ray binaries: Observation and theory}",
      journal = {New Astronomy Reviews},
         year = 2019,
        month = sep,
       volume = {85},
          eid = {101524},
        pages = {101524},
          doi = {10.1016/j.newar.2020.101524},
archivePrefix = {arXiv},
       eprint = {2001.08758},
 primaryClass = {astro-ph.HE},
       adsurl = {https://ui.adsabs.harvard.edu/abs/2019NewAR..8501524I}
}

@ARTICLE{Ingram2019a,
       author = {{Ingram}, Adam and {Mastroserio}, Guglielmo and {Dauser}, Thomas and {Hovenkamp}, Pieter and {van der Klis}, Michiel and {Garc{\'\i}a}, Javier A.},
        title = "{A public relativistic transfer function model for X-ray reverberation mapping of accreting black holes}",
      journal = {\mnras},
         year = 2019,
        month = sep,
       volume = {488},
       number = {1},
        pages = {324-347},
          doi = {10.1093/mnras/stz1720},
archivePrefix = {arXiv},
       eprint = {1906.08310},
 primaryClass = {astro-ph.HE},
       adsurl = {https://ui.adsabs.harvard.edu/abs/2019MNRAS.488..324I}
}

@ARTICLE{Mastroserio2021,
       author = {{Mastroserio}, Guglielmo and {Ingram}, Adam and {Wang}, Jingyi and {Garc{\'\i}a}, Javier A. and {van der Klis}, Michiel and {Cavecchi}, Yuri and {Connors}, Riley and {Dauser}, Thomas and {Harrison}, Fiona and {Kara}, Erin and {K{\"o}nig}, Ole and {Lucchini}, Matteo},
        title = "{Modelling correlated variability in accreting black holes: the effect of high density and variable ionization on reverberation lags}",
      journal = {\mnras},
         year = 2021,
        month = oct,
       volume = {507},
       number = {1},
        pages = {55-73},
          doi = {10.1093/mnras/stab2056},
archivePrefix = {arXiv},
       eprint = {2107.06893},
 primaryClass = {astro-ph.HE},
       adsurl = {https://ui.adsabs.harvard.edu/abs/2021MNRAS.507...55M}
}

@ARTICLE{Mastroserio2018,
       author = {{Mastroserio}, Guglielmo and {Ingram}, Adam and {van der Klis}, Michiel},
        title = "{Multi-time-scale X-ray reverberation mapping of accreting black holes}",
      journal = {\mnras},
         year = 2018,
        month = apr,
       volume = {475},
       number = {3},
        pages = {4027-4042},
          doi = {10.1093/mnras/sty075},
archivePrefix = {arXiv},
       eprint = {1801.03100},
 primaryClass = {astro-ph.HE},
       adsurl = {https://ui.adsabs.harvard.edu/abs/2018MNRAS.475.4027M}
}

@ARTICLE{Shreeram2020,
       author = {{Shreeram}, Soumya and {Ingram}, Adam},
        title = "{Exploring the radial disc ionization profile of the black hole X-ray binary GRS 1915+105}",
      journal = {\mnras},
         year = 2020,
        month = feb,
       volume = {492},
       number = {1},
        pages = {405-412},
          doi = {10.1093/mnras/stz3455},
archivePrefix = {arXiv},
       eprint = {1912.06833},
 primaryClass = {astro-ph.HE},
       adsurl = {https://ui.adsabs.harvard.edu/abs/2020MNRAS.492..405S}
}

@ARTICLE{Zdziarski1996,
       author = {{Zdziarski}, A.~A. and {Johnson}, W.~N. and {Magdziarz}, P.},
        title = "{Broad-band {\ensuremath{\gamma}}-ray and X-ray spectra of NGC 4151 and their implications for physical processes and geometry.}",
      journal = {\mnras},
         year = 1996,
        month = nov,
       volume = {283},
       number = {1},
        pages = {193-206},
          doi = {10.1093/mnras/283.1.193},
archivePrefix = {arXiv},
       eprint = {astro-ph/9607015},
 primaryClass = {astro-ph},
       adsurl = {https://ui.adsabs.harvard.edu/abs/1996MNRAS.283..193Z}
}

@ARTICLE{aaz2004,
       author = {{Zdziarski}, A.~A. and {Gierli{\'n}ski}, M.},
        title = "{Radiative Processes, Spectral States and Variability of Black-Hole Binaries}",
      journal = {Progress of Theoretical Physics Supplement},
         year = 2004,
        month = jan,
       volume = {155},
        pages = {99-119},
          doi = {10.1143/PTPS.155.99},
archivePrefix = {arXiv},
       eprint = {astro-ph/0403683},
 primaryClass = {astro-ph},
       adsurl = {https://ui.adsabs.harvard.edu/abs/2004PThPS.155...99Z}
}

@ARTICLE{nowak1999,
       author = {{Nowak}, Michael A. and {Vaughan}, Brian A. and {Wilms}, J{\"o}rn and {Dove}, James B. and {Begelman}, Mitchell C.},
        title = "{Rossi X-Ray Timing Explorer Observation of Cygnus X-1. II. Timing Analysis}",
      journal = {\apj},
         year = 1999,
        month = jan,
       volume = {510},
       number = {2},
        pages = {874-891},
          doi = {10.1086/306610},
archivePrefix = {arXiv},
       eprint = {astro-ph/9807278},
 primaryClass = {astro-ph},
       adsurl = {https://ui.adsabs.harvard.edu/abs/1999ApJ...510..874N}
}

@ARTICLE{demarco2021,
       author = {{De Marco}, B. and {Zdziarski}, A.~A. and {Ponti}, G. and {Migliori}, G. and {Belloni}, T.~M. and {Segovia Otero}, A. and {Dzie{\l}ak}, M.~A. and {Lai}, E.~V.},
        title = "{The inner flow geometry in MAXI J1820+070 during hard and hard-intermediate states}",
      journal = {\aap},
         year = 2021,
        month = oct,
       volume = {654},
          eid = {A14},
        pages = {A14},
          doi = {10.1051/0004-6361/202140567},
archivePrefix = {arXiv},
       eprint = {2102.07811},
 primaryClass = {astro-ph.HE},
       adsurl = {https://ui.adsabs.harvard.edu/abs/2021A&A...654A..14D}
}

@ARTICLE{kara2015,
       author = {{Kara}, E. and {Zoghbi}, A. and {Marinucci}, A. and {Walton}, D.~J. and {Fabian}, A.~C. and {Risaliti}, G. and {Boggs}, S.~E. and {Christensen}, F.~E. and {Fuerst}, F. and {Hailey}, C.~J. and {Harrison}, F.~A. and {Matt}, G. and {Parker}, M.~L. and {Reynolds}, C.~S. and {Stern}, D. and {Zhang}, W.~W.},
        title = "{Iron K and Compton hump reverberation in SWIFT J2127.4+5654 and NGC 1365 revealed by NuSTAR and XMM-Newton}",
      journal = {\mnras},
         year = 2015,
        month = jan,
       volume = {446},
       number = {1},
        pages = {737-749},
          doi = {10.1093/mnras/stu2136},
archivePrefix = {arXiv},
       eprint = {1410.3357},
 primaryClass = {astro-ph.HE},
       adsurl = {https://ui.adsabs.harvard.edu/abs/2015MNRAS.446..737K}
}

@ARTICLE{mizumoto2018,
       author = {{Mizumoto}, Misaki and {Done}, Chris and {Hagino}, Kouichi and {Ebisawa}, Ken and {Tsujimoto}, Masahiro and {Odaka}, Hirokazu},
        title = "{X-ray short-time lags in the Fe-K energy band produced by scattering clouds in active galactic nuclei}",
      journal = {\mnras},
         year = 2018,
        month = jul,
       volume = {478},
       number = {1},
        pages = {971-982},
          doi = {10.1093/mnras/sty1114},
archivePrefix = {arXiv},
       eprint = {1805.00046},
 primaryClass = {astro-ph.HE},
       adsurl = {https://ui.adsabs.harvard.edu/abs/2018MNRAS.478..971M}
}

@ARTICLE{wilkins2013,
       author = {{Wilkins}, D.~R. and {Fabian}, A.~C.},
        title = "{The origin of the lag spectra observed in AGN: Reverberation and the propagation of X-ray source fluctuations}",
      journal = {\mnras},
         year = 2013,
        month = mar,
       volume = {430},
       number = {1},
        pages = {247-258},
          doi = {10.1093/mnras/sts591},
archivePrefix = {arXiv},
       eprint = {1212.2213},
 primaryClass = {astro-ph.HE},
       adsurl = {https://ui.adsabs.harvard.edu/abs/2013MNRAS.430..247W}
}

@ARTICLE{uttley2002,
       author = {{Uttley}, P. and {McHardy}, I.~M. and {Papadakis}, I.~E.},
        title = "{Measuring the broad-band power spectra of active galactic nuclei with RXTE}",
      journal = {\mnras},
         year = 2002,
        month = may,
       volume = {332},
       number = {1},
        pages = {231-250},
          doi = {10.1046/j.1365-8711.2002.05298.x},
archivePrefix = {arXiv},
       eprint = {astro-ph/0201134},
 primaryClass = {astro-ph},
       adsurl = {https://ui.adsabs.harvard.edu/abs/2002MNRAS.332..231U}
}

@ARTICLE{demarco2015,
       author = {{De Marco}, B. and {Ponti}, G. and {Mu{\~n}oz-Darias}, T. and {Nandra}, K.},
        title = "{Tracing the Reverberation Lag in the Hard State of Black Hole X-Ray Binaries}",
      journal = {\apj},
         year = 2015,
        month = nov,
       volume = {814},
       number = {1},
          eid = {50},
        pages = {50},
          doi = {10.1088/0004-637X/814/1/50},
archivePrefix = {arXiv},
       eprint = {1510.02798},
 primaryClass = {astro-ph.HE},
       adsurl = {https://ui.adsabs.harvard.edu/abs/2015ApJ...814...50D}
}

@ARTICLE{zoghbi2010,
       author = {{Zoghbi}, A. and {Fabian}, A.~C. and {Uttley}, P. and {Miniutti}, G. and {Gallo}, L.~C. and {Reynolds}, C.~S. and {Miller}, J.~M. and {Ponti}, G.},
        title = "{Broad iron L line and X-ray reverberation in 1H0707-495}",
      journal = {\mnras},
         year = 2010,
        month = feb,
       volume = {401},
       number = {4},
        pages = {2419-2432},
          doi = {10.1111/j.1365-2966.2009.15816.x},
archivePrefix = {arXiv},
       eprint = {0910.0367},
 primaryClass = {astro-ph.HE},
       adsurl = {https://ui.adsabs.harvard.edu/abs/2010MNRAS.401.2419Z}
}

@ARTICLE{fabian2009,
       author = {{Fabian}, A.~C. and {Zoghbi}, A. and {Ross}, R.~R. and {Uttley}, P. and {Gallo}, L.~C. and {Brandt}, W.~N. and {Blustin}, A.~J. and {Boller}, T. and {Caballero-Garcia}, M.~D. and {Larsson}, J. and {Miller}, J.~M. and {Miniutti}, G. and {Ponti}, G. and {Reis}, R.~C. and {Reynolds}, C.~S. and {Tanaka}, Y. and {Young}, A.~J.},
        title = "{Broad line emission from iron K- and L-shell transitions in the active galaxy 1H0707-495}",
      journal = {\nat},
         year = 2009,
        month = may,
       volume = {459},
       number = {7246},
        pages = {540-542},
          doi = {10.1038/nature08007},
       adsurl = {https://ui.adsabs.harvard.edu/abs/2009Natur.459..540F}
}

@ARTICLE{garcia2013,
       author = {{Garc{\'\i}a}, J. and {Dauser}, T. and {Reynolds}, C.~S. and {Kallman}, T.~R. and {McClintock}, J.~E. and {Wilms}, J. and {Eikmann}, W.},
        title = "{X-Ray Reflected Spectra from Accretion Disk Models. III. A Complete Grid of Ionized Reflection Calculations}",
      journal = {\apj},
         year = 2013,
        month = may,
       volume = {768},
       number = {2},
          eid = {146},
        pages = {146},
          doi = {10.1088/0004-637X/768/2/146},
archivePrefix = {arXiv},
       eprint = {1303.2112},
 primaryClass = {astro-ph.HE},
       adsurl = {https://ui.adsabs.harvard.edu/abs/2013ApJ...768..146G}
}

@ARTICLE{uttley2011,
       author = {{Uttley}, P. and {Wilkinson}, T. and {Cassatella}, P. and {Wilms}, J. and {Pottschmidt}, K. and {Hanke}, M. and {B{\"o}ck}, M.},
        title = "{The causal connection between disc and power-law variability in hard state black hole X-ray binaries}",
      journal = {\mnras},
         year = 2011,
        month = jun,
       volume = {414},
       number = {1},
        pages = {L60-L64},
          doi = {10.1111/j.1745-3933.2011.01056.x},
archivePrefix = {arXiv},
       eprint = {1104.0634},
 primaryClass = {astro-ph.HE},
       adsurl = {https://ui.adsabs.harvard.edu/abs/2011MNRAS.414L..60U}
}

@ARTICLE{uttley2014,
       author = {{Uttley}, P. and {Cackett}, E.~M. and {Fabian}, A.~C. and {Kara}, E. and {Wilkins}, D.~R.},
        title = "{X-ray reverberation around accreting black holes}",
      journal = {\aapr},
         year = 2014,
        month = aug,
       volume = {22},
          eid = {72},
        pages = {72},
          doi = {10.1007/s00159-014-0072-0},
archivePrefix = {arXiv},
       eprint = {1405.6575},
 primaryClass = {astro-ph.HE},
       adsurl = {https://ui.adsabs.harvard.edu/abs/2014A&ARv..22...72U}
}

@ARTICLE{kara2019,
       author = {{Kara}, E. and {Steiner}, J.~F. and {Fabian}, A.~C. and {Cackett}, E.~M. and {Uttley}, P. and {Remillard}, R.~A. and {Gendreau}, K.~C. and {Arzoumanian}, Z. and {Altamirano}, D. and {Eikenberry}, S. and {Enoto}, T. and {Homan}, J. and {Neilsen}, J. and {Stevens}, A.~L.},
        title = "{The corona contracts in a black-hole transient}",
      journal = {\nat},
         year = 2019,
        month = jan,
       volume = {565},
       number = {7738},
        pages = {198-201},
          doi = {10.1038/s41586-018-0803-x},
archivePrefix = {arXiv},
       eprint = {1901.03877},
 primaryClass = {astro-ph.HE},
       adsurl = {https://ui.adsabs.harvard.edu/abs/2019Natur.565..198K}
}

@ARTICLE{vaughan2003,
       author = {{Vaughan}, S. and {Edelson}, R. and {Warwick}, R.~S. and {Uttley}, P.},
        title = "{On characterizing the variability properties of X-ray light curves from active galaxies}",
      journal = {\mnras},
         year = 2003,
        month = nov,
       volume = {345},
       number = {4},
        pages = {1271-1284},
          doi = {10.1046/j.1365-2966.2003.07042.x},
archivePrefix = {arXiv},
       eprint = {astro-ph/0307420},
 primaryClass = {astro-ph},
       adsurl = {https://ui.adsabs.harvard.edu/abs/2003MNRAS.345.1271V}
}

@INPROCEEDINGS{zhang2014,
       author = {{Zhang}, S. and {Lu}, F.~J. and {Zhang}, S.~N. and {Li}, T.~P.},
        title = "{Introduction to the hard x-ray modulation telescope}",
    booktitle = {Space Telescopes and Instrumentation 2014: Ultraviolet to Gamma Ray},
         year = 2014,
       editor = {{Takahashi}, Tadayuki and {den Herder}, Jan-Willem A. and {Bautz}, Mark},
       series = {Society of Photo-Optical Instrumentation Engineers (SPIE) Conference Series},
       volume = {9144},
        month = jul,
          eid = {914421},
        pages = {914421},
          doi = {10.1117/12.2054144},
       adsurl = {https://ui.adsabs.harvard.edu/abs/2014SPIE.9144E..21Z}
}

@article{cassatella2012joint,
  title={Joint spectral-timing modelling of the hard lags in GX 339- 4: constraints on reflection models},
  author={Cassatella, Pablo and Uttley, Phil and Wilms, Joern and Poutanen, Juri},
  journal={Monthly Notices of the Royal Astronomical Society},
  volume={422},
  number={3},
  pages={2407--2416},
  year={2012},
  publisher={The Royal Astronomical Society}
}

@article{de2017evolution,
  title={Evolution of the reverberation lag in GX 339--4 at the end of an outburst},
  author={De Marco, B and Ponti, GABRIELE and Petrucci, PO and Clavel, M and Corbel, St{\'e}phane and Belmont, R and Chakravorty, S and Coriat, M and Drappeau, S and Ferreira, Jorge and others},
  journal={Monthly Notices of the Royal Astronomical Society},
  volume={471},
  number={2},
  pages={1475--1487},
  year={2017},
  publisher={Oxford University Press}
}

@ARTICLE{2001MNRAS.327..799K,
       author = {{Kotov}, O. and {Churazov}, E. and {Gilfanov}, M.},
        title = "{On the X-ray time-lags in the black hole candidates}",
      journal = {\mnras},
         year = 2001,
        month = nov,
       volume = {327},
       number = {3},
        pages = {799-807},
          doi = {10.1046/j.1365-8711.2001.04769.x},
archivePrefix = {arXiv},
       eprint = {astro-ph/0103115},
 primaryClass = {astro-ph},
       adsurl = {https://ui.adsabs.harvard.edu/abs/2001MNRAS.327..799K}
}

@ARTICLE{2006MNRAS.367..801A,
       author = {{Ar{\'e}valo}, P. and {Uttley}, P.},
        title = "{Investigating a fluctuating-accretion model for the spectral-timing properties of accreting black hole systems}",
      journal = {\mnras},
         year = 2006,
        month = apr,
       volume = {367},
       number = {2},
        pages = {801-814},
          doi = {10.1111/j.1365-2966.2006.09989.x},
archivePrefix = {arXiv},
       eprint = {astro-ph/0512394},
 primaryClass = {astro-ph},
       adsurl = {https://ui.adsabs.harvard.edu/abs/2006MNRAS.367..801A}
}

@ARTICLE{1997MNRAS.292..679L,
       author = {{Lyubarskii}, Yu. E.},
        title = "{Flicker noise in accretion discs}",
      journal = {\mnras},
         year = 1997,
        month = dec,
       volume = {292},
       number = {3},
        pages = {679-685},
          doi = {10.1093/mnras/292.3.679},
       adsurl = {https://ui.adsabs.harvard.edu/abs/1997MNRAS.292..679L}
}

@ARTICLE{2001MNRAS.321..759C,
       author = {{Churazov}, E. and {Gilfanov}, M. and {Revnivtsev}, M.},
        title = "{Soft state of Cygnus X-1: stable disc and unstable corona}",
      journal = {\mnras},
         year = 2001,
        month = mar,
       volume = {321},
       number = {4},
        pages = {759-766},
          doi = {10.1046/j.1365-8711.2001.04056.x},
archivePrefix = {arXiv},
       eprint = {astro-ph/0006227},
 primaryClass = {astro-ph},
       adsurl = {https://ui.adsabs.harvard.edu/abs/2001MNRAS.321..759C}
}

@ARTICLE{2012MNRAS.419.2369I,
       author = {{Ingram}, Adam and {Done}, Chris},
        title = "{Modelling variability in black hole binaries: linking simulations to observations}",
      journal = {\mnras},
         year = 2012,
        month = jan,
       volume = {419},
       number = {3},
        pages = {2369-2378},
          doi = {10.1111/j.1365-2966.2011.19885.x},
archivePrefix = {arXiv},
       eprint = {1108.0789},
 primaryClass = {astro-ph.HE},
       adsurl = {https://ui.adsabs.harvard.edu/abs/2012MNRAS.419.2369I}
}

@ARTICLE{2011MNRAS.415.2323I,
       author = {{Ingram}, Adam and {Done}, Chris},
        title = "{A physical model for the continuum variability and quasi-periodic oscillation in accreting black holes}",
      journal = {\mnras},
         year = 2011,
        month = aug,
       volume = {415},
       number = {3},
        pages = {2323-2335},
          doi = {10.1111/j.1365-2966.2011.18860.x},
archivePrefix = {arXiv},
       eprint = {1101.2336},
 primaryClass = {astro-ph.SR},
       adsurl = {https://ui.adsabs.harvard.edu/abs/2011MNRAS.415.2323I}
}

@ARTICLE{2013MNRAS.434.1476I,
       author = {{Ingram}, Adam and {van der Klis}, Michiel},
        title = "{An exact analytic treatment of propagating mass accretion rate fluctuations in X-ray binaries}",
      journal = {\mnras},
         year = 2013,
        month = sep,
       volume = {434},
       number = {2},
        pages = {1476-1485},
          doi = {10.1093/mnras/stt1107},
archivePrefix = {arXiv},
       eprint = {1306.3823},
 primaryClass = {astro-ph.HE},
       adsurl = {https://ui.adsabs.harvard.edu/abs/2013MNRAS.434.1476I}
}

@ARTICLE{2014MNRAS.440.2882R,
       author = {{Rapisarda}, S. and {Ingram}, A. and {van der Klis}, M.},
        title = "{Evolution of the hot flow of MAXI J1543-564}",
      journal = {\mnras},
         year = 2014,
        month = may,
       volume = {440},
       number = {3},
        pages = {2882-2893},
          doi = {10.1093/mnras/stu461},
archivePrefix = {arXiv},
       eprint = {1403.2308},
 primaryClass = {astro-ph.HE},
       adsurl = {https://ui.adsabs.harvard.edu/abs/2014MNRAS.440.2882R}
}

@ARTICLE{2016MNRAS.462.4078R,
       author = {{Rapisarda}, S. and {Ingram}, A. and {Kalamkar}, M. and {van der Klis}, M.},
        title = "{Modelling the cross-spectral variability of the black hole binary MAXI J1659-152 with propagating accretion rate fluctuations}",
      journal = {\mnras},
         year = 2016,
        month = nov,
       volume = {462},
       number = {4},
        pages = {4078-4093},
          doi = {10.1093/mnras/stw1878},
archivePrefix = {arXiv},
       eprint = {1607.08178},
 primaryClass = {astro-ph.HE},
       adsurl = {https://ui.adsabs.harvard.edu/abs/2016MNRAS.462.4078R}
}

@ARTICLE{1999ApJ...510..874N,
       author = {{Nowak}, Michael A. and {Vaughan}, Brian A. and {Wilms}, J{\"o}rn and {Dove}, James B. and {Begelman}, Mitchell C.},
        title = "{Rossi X-Ray Timing Explorer Observation of Cygnus X-1. II. Timing Analysis}",
      journal = {\apj},
         year = 1999,
        month = jan,
       volume = {510},
       number = {2},
        pages = {874-891},
          doi = {10.1086/306610},
archivePrefix = {arXiv},
       eprint = {astro-ph/9807278},
 primaryClass = {astro-ph},
       adsurl = {https://ui.adsabs.harvard.edu/abs/1999ApJ...510..874N}
}

@ARTICLE{2000A&A...357L..17P,
       author = {{Pottschmidt}, K. and {Wilms}, J. and {Nowak}, M.~A. and {Heindl}, W.~A. and {Smith}, D.~M. and {Staubert}, R.},
        title = "{Temporal evolution of X-ray lags in Cygnus X-1}",
      journal = {\aap},
         year = 2000,
        month = may,
       volume = {357},
        pages = {L17-L20},
          doi = {10.48550/arXiv.astro-ph/0004018},
archivePrefix = {arXiv},
       eprint = {astro-ph/0004018},
 primaryClass = {astro-ph},
       adsurl = {https://ui.adsabs.harvard.edu/abs/2000A&A...357L..17P}
}

@ARTICLE{Svoboda2012,
       author = {{Svoboda}, J. and {Dov{\v{c}}iak}, M. and {Goosmann}, R.~W. and {Jethwa}, P. and {Karas}, V. and {Miniutti}, G. and {Guainazzi}, M.},
        title = "{Origin of the X-ray disc-reflection steep radial emissivity}",
      journal = {\aap},
         year = 2012,
        month = sep,
       volume = {545},
          eid = {A106},
        pages = {A106},
          doi = {10.1051/0004-6361/201219701},
archivePrefix = {arXiv},
       eprint = {1208.0360},
 primaryClass = {astro-ph.HE},
       adsurl = {https://ui.adsabs.harvard.edu/abs/2012A&A...545A.106S}
}

@ARTICLE{Shakura1973,
       author = {{Shakura}, N.~I. and {Sunyaev}, R.~A.},
        title = "{Black holes in binary systems. Observational appearance.}",
      journal = {\aap},
         year = 1973,
        month = jan,
       volume = {24},
        pages = {337-355},
       adsurl = {https://ui.adsabs.harvard.edu/abs/1973A&A....24..337S}
}

@ARTICLE{athena,
       author = {{Nandra}, Kirpal and {Barret}, Didier and {Barcons}, Xavier and {Fabian}, Andy and {den Herder}, Jan-Willem and {Piro}, Luigi and {Watson}, Mike and {Adami}, Christophe and {Aird}, James and {Afonso}, Jose Manuel and {Alexander}, Dave and {Argiroffi}, Costanza and {Amati}, Lorenzo and {Arnaud}, Monique and {Atteia}, Jean-Luc and {Audard}, Marc and {Badenes}, Carles and {Ballet}, Jean and {Ballo}, Lucia and {Bamba}, Aya and {Bhardwaj}, Anil and {Stefano Battistelli}, Elia and {Becker}, Werner and {De Becker}, Micha{\"e}l and {Behar}, Ehud and {Bianchi}, Stefano and {Biffi}, Veronica and {B{\^\i}rzan}, Laura and {Bocchino}, Fabrizio and {Bogdanov}, Slavko and {Boirin}, Laurence and {Boller}, Thomas and {Borgani}, Stefano and {Borm}, Katharina and {Bouch{\'e}}, Nicolas and {Bourdin}, Herv{\'e} and {Bower}, Richard and {Braito}, Valentina and {Branchini}, Enzo and {Branduardi-Raymont}, Graziella and {Bregman}, Joel and {Brenneman}, Laura and {Brightman}, Murray and {Br{\"u}ggen}, Marcus and {Buchner}, Johannes and {Bulbul}, Esra and {Brusa}, Marcella and {Bursa}, Michal and {Caccianiga}, Alessandro and {Cackett}, Ed and {Campana}, Sergio and {Cappelluti}, Nico and {Cappi}, Massimo and {Carrera}, Francisco and {Ceballos}, Maite and {Christensen}, Finn and {Chu}, You-Hua and {Churazov}, Eugene and {Clerc}, Nicolas and {Corbel}, Stephane and {Corral}, Amalia and {Comastri}, Andrea and {Costantini}, Elisa and {Croston}, Judith and {Dadina}, Mauro and {D'Ai}, Antonino and {Decourchelle}, Anne and {Della Ceca}, Roberto and {Dennerl}, Konrad and {Dolag}, Klaus and {Done}, Chris and {Dovciak}, Michal and {Drake}, Jeremy and {Eckert}, Dominique and {Edge}, Alastair and {Ettori}, Stefano and {Ezoe}, Yuichiro and {Feigelson}, Eric and {Fender}, Rob and {Feruglio}, Chiara and {Finoguenov}, Alexis and {Fiore}, Fabrizio and {Galeazzi}, Massimiliano and {Gallagher}, Sarah and {Gandhi}, Poshak and {Gaspari}, Massimo and {Gastaldello}, Fabio and {Georgakakis}, Antonis and {Georgantopoulos}, Ioannis and {Gilfanov}, Marat and {Gitti}, Myriam and {Gladstone}, Randy and {Goosmann}, Rene and {Gosset}, Eric and {Grosso}, Nicolas and {Guedel}, Manuel and {Guerrero}, Martin and {Haberl}, Frank and {Hardcastle}, Martin and {Heinz}, Sebastian and {Alonso Herrero}, Almudena and {Herv{\'e}}, Anthony and {Holmstrom}, Mats and {Iwasawa}, Kazushi and {Jonker}, Peter and {Kaastra}, Jelle and {Kara}, Erin and {Karas}, Vladimir and {Kastner}, Joel and {King}, Andrew and {Kosenko}, Daria and {Koutroumpa}, Dimita and {Kraft}, Ralph and {Kreykenbohm}, Ingo and {Lallement}, Rosine and {Lanzuisi}, Giorgio and {Lee}, J. and {Lemoine-Goumard}, Marianne and {Lobban}, Andrew and {Lodato}, Giuseppe and {Lovisari}, Lorenzo and {Lotti}, Simone and {McCharthy}, Ian and {McNamara}, Brian and {Maggio}, Antonio and {Maiolino}, Roberto and {De Marco}, Barbara and {de Martino}, Domitilla and {Mateos}, Silvia and {Matt}, Giorgio and {Maughan}, Ben and {Mazzotta}, Pasquale and {Mendez}, Mariano and {Merloni}, Andrea and {Micela}, Giuseppina and {Miceli}, Marco and {Mignani}, Robert and {Miller}, Jon and {Miniutti}, Giovanni and {Molendi}, Silvano and {Montez}, Rodolfo and {Moretti}, Alberto and {Motch}, Christian and {Naz{\'e}}, Ya{\"e}l and {Nevalainen}, Jukka and {Nicastro}, Fabrizio and {Nulsen}, Paul and {Ohashi}, Takaya and {O'Brien}, Paul and {Osborne}, Julian and {Oskinova}, Lida and {Pacaud}, Florian and {Paerels}, Frederik and {Page}, Mat and {Papadakis}, Iossif and {Pareschi}, Giovanni and {Petre}, Robert and {Petrucci}, Pierre-Olivier and {Piconcelli}, Enrico and {Pillitteri}, Ignazio and {Pinto}, C. and {de Plaa}, Jelle and {Pointecouteau}, Etienne and {Ponman}, Trevor and {Ponti}, Gabriele and {Porquet}, Delphine and {Pounds}, Ken and {Pratt}, Gabriel and {Predehl}, Peter and {Proga}, Daniel and {Psaltis}, Dimitrios and {Rafferty}, David and {Ramos-Ceja}, Miriam and {Ranalli}, Piero and {Rasia}, Elena and {Rau}, Arne and {Rauw}, Gregor and {Rea}, Nanda and {Read}, Andy and {Reeves}, James and {Reiprich}, Thomas and {Renaud}, Matthieu and {Reynolds}, Chris and {Risaliti}, Guido and {Rodriguez}, Jerome and {Rodriguez Hidalgo}, Paola and {Roncarelli}, Mauro and {Rosario}, David and {Rossetti}, Mariachiara and {Rozanska}, Agata and {Rovilos}, Emmanouil and {Salvaterra}, Ruben and {Salvato}, Mara and {Di Salvo}, Tiziana and {Sanders}, Jeremy and {Sanz-Forcada}, Jorge and {Schawinski}, Kevin and {Schaye}, Joop and {Schwope}, Axel and {Sciortino}, Salvatore},
        title = "{The Hot and Energetic Universe: A White Paper presenting the science theme motivating the Athena+ mission}",
      journal = {arXiv e-prints},
         year = 2013,
        month = jun,
          eid = {arXiv:1306.2307},
        pages = {arXiv:1306.2307},
          doi = {10.48550/arXiv.1306.2307},
archivePrefix = {arXiv},
       eprint = {1306.2307},
 primaryClass = {astro-ph.HE},
       adsurl = {https://ui.adsabs.harvard.edu/abs/2013arXiv1306.2307N}
}

@ARTICLE{extp,
       author = {{De Rosa}, Alessandra and {Uttley}, Phil and {Gou}, LiJun and {Liu}, Yuan and {Bambi}, Cosimo and {Barret}, Didier and {Belloni}, Tomaso and {Berti}, Emanuele and {Bianchi}, Stefano and {Caiazzo}, Ilaria and {Casella}, Piergiorgio and {Feroci}, Marco and {Ferrari}, Valeria and {Gualtieri}, Leonardo and {Heyl}, Jeremy and {Ingram}, Adam and {Karas}, Vladimir and {Lu}, FangJun and {Luo}, Bin and {Matt}, Giorgio and {Motta}, Sara and {Neilsen}, Joseph and {Pani}, Paolo and {Santangelo}, Andrea and {Shu}, XinWen and {Wang}, JunFeng and {Wang}, Jian-Min and {Xue}, YongQuan and {Xu}, YuPeng and {Yuan}, WeiMin and {Yuan}, YeFei and {Zhang}, Shuang-Nan and {Zhang}, Shu and {Agudo}, Ivan and {Amati}, Lorenzo and {Andersson}, Nils and {Baglio}, Cristina and {Bakala}, Pavel and {Baykal}, Altan and {Bhattacharyya}, Sudip and {Bombaci}, Ignazio and {Bucciantini}, Niccol{\'o} and {Capitanio}, Fiamma and {Ciolfi}, Riccardo and {Cui}, Wei K. and {D'Ammando}, Filippo and {Dauser}, Thomas and {Del Santo}, Melania and {De Marco}, Barbara and {Di Salvo}, Tiziana and {Done}, Chris and {Dov{\v{c}}iak}, Michal and {Fabian}, Andrew C. and {Falanga}, Maurizio and {Gambino}, Angelo Francesco and {Gendre}, Bruce and {Grinberg}, Victoria and {Heger}, Alexander and {Homan}, Jeroen and {Iaria}, Rosario and {Jiang}, JiaChen and {Jin}, ChiChuan and {Koerding}, Elmar and {Linares}, Manu and {Liu}, Zhu and {Maccarone}, Thomas J. and {Malzac}, Julien and {Manousakis}, Antonios and {Marin}, Fr{\'e}d{\'e}ric and {Marinucci}, Andrea and {Mehdipour}, Missagh and {M{\'e}ndez}, Mariano and {Migliari}, Simone and {Miller}, Cole and {Miniutti}, Giovanni and {Nardini}, Emanuele and {O'Brien}, Paul T. and {Osborne}, Julian P. and {Petrucci}, Pierre Olivier and {Possenti}, Andrea and {Riggio}, Alessandro and {Rodriguez}, Jerome and {Sanna}, Andrea and {Shao}, LiJing and {Sobolewska}, Malgosia and {Sramkova}, Eva and {Stevens}, Abigail L. and {Stiele}, Holger and {Stratta}, Giulia and {Stuchlik}, Zdenek and {Svoboda}, Jiri and {Tamburini}, Fabrizio and {Tauris}, Thomas M. and {Tombesi}, Francesco and {Torok}, Gabriel and {Urbanec}, Martin and {Vincent}, Frederic and {Wu}, QingWen and {Yuan}, Feng and {in't Zand}, Jean J.~M. and {Zdziarski}, Andrzej A. and {Zhou}, XinLin},
        title = "{Accretion in strong field gravity with eXTP}",
      journal = {Science China Physics, Mechanics, and Astronomy},
         year = 2019,
        month = feb,
       volume = {62},
       number = {2},
          eid = {29504},
        pages = {29504},
          doi = {10.1007/s11433-018-9297-0},
archivePrefix = {arXiv},
       eprint = {1812.04022},
 primaryClass = {astro-ph.HE},
       adsurl = {https://ui.adsabs.harvard.edu/abs/2019SCPMA..6229504D}
}

@INPROCEEDINGS{nicer,
       author = {{Gendreau}, Keith C. and {Arzoumanian}, Zaven and {Adkins}, Phillip W. and {Albert}, Cheryl L. and {Anders}, John F. and {Aylward}, Andrew T. and {Baker}, Charles L. and {Balsamo}, Erin R. and {Bamford}, William A. and {Benegalrao}, Suyog S. and {Berry}, Daniel L. and {Bhalwani}, Shiraz and {Black}, J. Kevin and {Blaurock}, Carl and {Bronke}, Ginger M. and {Brown}, Gary L. and {Budinoff}, Jason G. and {Cantwell}, Jeffrey D. and {Cazeau}, Thoniel and {Chen}, Philip T. and {Clement}, Thomas G. and {Colangelo}, Andrew T. and {Coleman}, Jerry S. and {Coopersmith}, Jonathan D. and {Dehaven}, William E. and {Doty}, John P. and {Egan}, Mark D. and {Enoto}, Teruaki and {Fan}, Terry W. and {Ferro}, Deneen M. and {Foster}, Richard and {Galassi}, Nicholas M. and {Gallo}, Luis D. and {Green}, Chris M. and {Grosh}, Dave and {Ha}, Kong Q. and {Hasouneh}, Monther A. and {Heefner}, Kristofer B. and {Hestnes}, Phyllis and {Hoge}, Lisa J. and {Jacobs}, Tawanda M. and {J{\o}rgensen}, John L. and {Kaiser}, Michael A. and {Kellogg}, James W. and {Kenyon}, Steven J. and {Koenecke}, Richard G. and {Kozon}, Robert P. and {LaMarr}, Beverly and {Lambertson}, Mike D. and {Larson}, Anne M. and {Lentine}, Steven and {Lewis}, Jesse H. and {Lilly}, Michael G. and {Liu}, Kuochia Alice and {Malonis}, Andrew and {Manthripragada}, Sridhar S. and {Markwardt}, Craig B. and {Matonak}, Bryan D. and {Mcginnis}, Isaac E. and {Miller}, Roger L. and {Mitchell}, Alissa L. and {Mitchell}, Jason W. and {Mohammed}, Jelila S. and {Monroe}, Charles A. and {Montt de Garcia}, Kristina M. and {Mul{\'e}}, Peter D. and {Nagao}, Louis T. and {Ngo}, Son N. and {Norris}, Eric D. and {Norwood}, Dwight A. and {Novotka}, Joseph and {Okajima}, Takashi and {Olsen}, Lawrence G. and {Onyeachu}, Chimaobi O. and {Orosco}, Henry Y. and {Peterson}, Jacqualine R. and {Pevear}, Kristina N. and {Pham}, Karen K. and {Pollard}, Sue E. and {Pope}, John S. and {Powers}, Daniel F. and {Powers}, Charles E. and {Price}, Samuel R. and {Prigozhin}, Gregory Y. and {Ramirez}, Julian B. and {Reid}, Winston J. and {Remillard}, Ronald A. and {Rogstad}, Eric M. and {Rosecrans}, Glenn P. and {Rowe}, John N. and {Sager}, Jennifer A. and {Sanders}, Claude A. and {Savadkin}, Bruce and {Saylor}, Maxine R. and {Schaeffer}, Alexander F. and {Schweiss}, Nancy S. and {Semper}, Sean R. and {Serlemitsos}, Peter J. and {Shackelford}, Larry V. and {Soong}, Yang and {Struebel}, Jonathan and {Vezie}, Michael L. and {Villasenor}, Joel S. and {Winternitz}, Luke B. and {Wofford}, George I. and {Wright}, Michael R. and {Yang}, Mike Y. and {Yu}, Wayne H.},
        title = "{The Neutron star Interior Composition Explorer (NICER): design and development}",
    booktitle = {Space Telescopes and Instrumentation 2016: Ultraviolet to Gamma Ray},
         year = 2016,
       editor = {{den Herder}, Jan-Willem A. and {Takahashi}, Tadayuki and {Bautz}, Marshall},
       series = {Society of Photo-Optical Instrumentation Engineers (SPIE) Conference Series},
       volume = {9905},
        month = jul,
          eid = {99051H},
        pages = {99051H},
          doi = {10.1117/12.2231304},
       adsurl = {https://ui.adsabs.harvard.edu/abs/2016SPIE.9905E..1HG}
}

@ARTICLE{2022ApJ...930...18W,
       author = {{Wang}, Jingyi and {Kara}, Erin and {Lucchini}, Matteo and {Ingram}, Adam and {van der Klis}, Michiel and {Mastroserio}, Guglielmo and {Garc{\'\i}a}, Javier A. and {Dauser}, Thomas and {Connors}, Riley and {Fabian}, Andrew C. and {Steiner}, James F. and {Remillard}, Ron A. and {Cackett}, Edward M. and {Uttley}, Phil and {Altamirano}, Diego},
        title = "{The NICER ``Reverberation Machine'': A Systematic Study of Time Lags in Black Hole X-Ray Binaries}",
      journal = {\apj},
         year = 2022,
        month = may,
       volume = {930},
       number = {1},
          eid = {18},
        pages = {18},
          doi = {10.3847/1538-4357/ac6262},
archivePrefix = {arXiv},
       eprint = {2205.00928},
 primaryClass = {astro-ph.HE},
       adsurl = {https://ui.adsabs.harvard.edu/abs/2022ApJ...930...18W}
}

@article{zoghbi2014observations,
  title={OBSERVATIONS OF MCG--5-23-16 WITH SUZAKU, XMM-NEWTON AND NUSTAR: DISK TOMOGRAPHY AND COMPTON HUMP REVERBERATION},
  author={Zoghbi, A and Cackett, EM and Reynolds, C and Kara, E and Harrison, FA and Fabian, AC and Lohfink, A and Matt, G and Balokovic, M and Boggs, SE and others},
  journal={The Astrophysical Journal},
  volume={789},
  number={1},
  pages={56},
  year={2014},
  publisher={IOP Publishing}
}

@article{lyubarskii1997flicker,
  title={Flicker noise in accretion discs},
  author={Lyubarskii, Yu E},
  journal={Monthly Notices of the Royal Astronomical Society},
  volume={292},
  number={3},
  pages={679--685},
  year={1997},
  publisher={Blackwell Science Ltd Oxford, UK}
}

@article{wilkinson2009accretion,
  title={Accretion disc variability in the hard state of black hole X-ray binaries},
  author={Wilkinson, Tony and Uttley, Philip},
  journal={Monthly Notices of the Royal Astronomical Society},
  volume={397},
  number={2},
  pages={666--676},
  year={2009},
  publisher={Blackwell Publishing Ltd Oxford, UK}
}

@article{balbus1998instability,
  title={Instability, turbulence, and enhanced transport in accretion disks},
  author={Balbus, Steven A and Hawley, John F},
  journal={Reviews of modern physics},
  volume={70},
  number={1},
  pages={1},
  year={1998},
  publisher={APS}
}

@article{fabian2010x,
  title={X-ray Reflection},
  author={Fabian, AC and Ross, RR},
  journal={Space science reviews},
  volume={157},
  pages={167--176},
  year={2010},
  publisher={Springer}
}

@article{dauser2016relativistic,
  title={Relativistic reflection: Review and recent developments in modeling},
  author={Dauser, Thomas and Garc{\'\i}a, J and Wilms, J{\"o}rn},
  journal={Astronomische Nachrichten},
  volume={337},
  number={4-5},
  pages={362--367},
  year={2016},
  publisher={Wiley Online Library}
}

@article{mchardy2006active,
  title={Active galactic nuclei as scaled-up Galactic black holes},
  author={McHardy, Ian M and Koerding, E and Knigge, C and Uttley, P and Fender, RP},
  journal={Nature},
  volume={444},
  number={7120},
  pages={730--732},
  year={2006},
  publisher={Nature Publishing Group UK London}
}

@article{arevalo2006spectral,
  title={Spectral-timing evidence for a very high state in the narrow-line Seyfert 1 Ark 564},
  author={Ar{\'e}valo, Patricia and Papadakis, IE and Uttley, P and McHardy, IM and Brinkmann, W},
  journal={Monthly Notices of the Royal Astronomical Society},
  volume={372},
  number={1},
  pages={401--409},
  year={2006},
  publisher={Blackwell Publishing Ltd Oxford, UK}
}

@article{mendez2022coupling,
  title={Coupling between the accreting corona and the relativistic jet in the microquasar GRS 1915+ 105},
  author={M{\'e}ndez, Mariano and Karpouzas, Konstantinos and Garc{\'\i}a, Federico and Zhang, Liang and Zhang, Yuexin and Belloni, Tomaso M and Altamirano, Diego},
  journal={Nature Astronomy},
  volume={6},
  number={5},
  pages={577--583},
  year={2022},
  publisher={Nature Publishing Group UK London}
}

@article{wandel1986observational,
  title={Observational determination of the masses of active galactic nuclei},
  author={Wandel, Amri and Mushotzky, Richard Fred},
  journal={Astrophysical Journal, Part 2-Letters to the Editor (ISSN 0004-637X), vol. 306, July 15, 1986, p. L61-L65.},
  volume={306},
  pages={L61--L65},
  year={1986}
}

@article{ponti2012caixa,
  title={CAIXA: a catalogue of AGN in the XMM-Newton archive-III. Excess variance analysis},
  author={Ponti, Gabriele and Papadakis, Iossif and Bianchi, Stefano and Guainazzi, Matteo and Matt, Giorgio and Uttley, Phil and Bonilla, NF},
  journal={Astronomy \& Astrophysics},
  volume={542},
  pages={A83},
  year={2012},
  publisher={EDP Sciences}
}

@article{malizia2008first,
  title={First high-energy observations of narrow-line Seyfert 1s with INTEGRAL/IBIS},
  author={Malizia, A and Bassani, L and Bird, AJ and Landi, R and Masetti, N and De Rosa, A and Panessa, F and Molina, M and Dean, AJ and Perri, M and others},
  journal={Monthly Notices of the Royal Astronomical Society},
  volume={389},
  number={3},
  pages={1360--1366},
  year={2008},
  publisher={Blackwell Publishing Ltd Oxford, UK}
}

@article{marconi2003relation,
  title={The relation between black hole mass, bulge mass, and near-infrared luminosity},
  author={Marconi, Alessandro and Hunt, Leslie K},
  journal={The Astrophysical Journal},
  volume={589},
  number={1},
  pages={L21},
  year={2003},
  publisher={IOP Publishing}
}

@article{emmanoulopoulos2014general,
  title={General relativistic modelling of the negative reverberation X-ray time delays in AGN},
  author={Emmanoulopoulos, D and Papadakis, IE and Dov{\v{c}}iak, M and McHardy, IM},
  journal={Monthly Notices of the Royal Astronomical Society},
  volume={439},
  number={4},
  pages={3931--3950},
  year={2014},
  publisher={Oxford Academic}
}

@article{kylafis2008jet,
  title={A jet model for Galactic black-hole X-ray sources: some constraining correlations},
  author={Kylafis, ND and Papadakis, IE and Reig, P and Giannios, D and Pooley, GG},
  journal={Astronomy \& Astrophysics},
  volume={489},
  number={2},
  pages={481--487},
  year={2008},
  publisher={EDP Sciences}
}

@ARTICLE{2022NatAs...6..577M,
       author = {{M{\'e}ndez}, Mariano and {Karpouzas}, Konstantinos and {Garc{\'\i}a}, Federico and {Zhang}, Liang and {Zhang}, Yuexin and {Belloni}, Tomaso M. and {Altamirano}, Diego},
        title = "{Coupling between the accreting corona and the relativistic jet in the microquasar GRS 1915+105}",
      journal = {Nature Astronomy},
         year = 2022,
        month = mar,
       volume = {6},
        pages = {577-583},
          doi = {10.1038/s41550-022-01617-y},
archivePrefix = {arXiv},
       eprint = {2203.02963},
 primaryClass = {astro-ph.HE},
       adsurl = {https://ui.adsabs.harvard.edu/abs/2022NatAs...6..577M}
}

@article{uttley2025large,
  title={Large and complex X-ray time lags from black hole accretion discs with compact inner coronae},
  author={Uttley, Phil and Malzac, Julien},
  journal={Monthly Notices of the Royal Astronomical Society},
  volume={536},
  number={4},
  pages={3284--3307},
  year={2025},
  publisher={Oxford University Press}
}

@article{mahmoud2019reverberation,
  title={Reverberation reveals the truncated disc in the hard state of GX 339-4},
  author={Mahmoud, Ra’ad D and Done, Chris and De Marco, Barbara},
  journal={Monthly Notices of the Royal Astronomical Society},
  volume={486},
  number={2},
  pages={2137--2152},
  year={2019},
  publisher={Oxford University Press}
}

@article{wang2020evolution,
  title={The evolution of the broadband temporal features observed in the black-hole transient MAXI J1820+ 070 with insight-HXMT},
  author={Wang, Yanan and Ji, Long and Zhang, SN and M{\'e}ndez, Mariano and Qu, JL and Maggi, Pierre and Ge, MY and Qiao, Erlin and Tao, L and Zhang, S and others},
  journal={The Astrophysical Journal},
  volume={896},
  number={1},
  pages={33},
  year={2020},
  publisher={IoP Publishing}
}

@article{kawamura2023maxi,
  title={MAXI J1820+ 070 X-ray spectral-timing reveals the nature of the accretion flow in black hole binaries},
  author={Kawamura, Tenyo and Done, Chris and Axelsson, Magnus and Takahashi, Tadayuki},
  journal={Monthly Notices of the Royal Astronomical Society},
  volume={519},
  number={3},
  pages={4434--4453},
  year={2023},
  publisher={Oxford University Press}
}


\end{document}